\documentclass[12pt]{article}

\usepackage{pdfpages}
\pdfoutput=1

\usepackage{amsfonts}

\def\ir{I\kern-.1667em R\ }
\newcommand{\BEQ}{\begin{equation}}
\newcommand{\NEQ}{\end{equation}}
\def\nl{\hfill\break}       
\def\np{\vfill\eject}       
\def\nsect{\vskip 2pc\noindent}       
\def\ni{\noindent}
\def\IR{I\kern-.255em R}

\def\hal{\hat\alpha}
\def\hsi{\hat\sigma}
\def\hsig{\hat\sigma}   

\def\Ze{Z_{eff} }

\def\sigs{\hsig^2}

\def\s2trl{\hsig_{tot}^2(r_l) }

\def\eps{\epsilon }
\def\nb{\overline{n_e} }
\def\conp{ $ \nb,\ I_p$  and $q_{ a }$ }
\def\conpr{ $ \qc,\ I_p$  and $\nb    $ }

\def\equil{mhd}
\def\Eequil{Emhd}
\def\wmhd{{\rm Wp}_{mhd}}
\def\wdia{{\rm Wp}_{dia}}
\def\wkin{{\rm Wp}_{kin}}

\def \qc{q_{ a }}

\def\xv{ \vec{x}}

\def\Yu{\underline Y\,}
\def\Yb{\overline{Y} }

\def\Siuu{\underline{\underline \Sigma}\, }

\def\Siuuhat{\underline{\underline {\hat\Sigma}}\, }

\def\Si{\Sigma}
\def\Sihat{\hat\Sigma}

\def\yh{\hat y}
\def\alu{\underline\alpha}
\def\thu{\underline\theta}

\def\bpd{\beta_{p(dia)}}
\def\Rs{${\rm R^2} $}
\def\rt{ r_{trans} }

\def\bYA{Y\kern-0.73em Y}
\def\bya{y\kern-0.57em y}
\def\bYB{Y\kern-0.74em Y}
\def\byb{y\kern-0.58em y}
\def\bYC{Y\kern-0.75em Y}
\def\byc{y\kern-0.59em y}

\def\frac#1#2{{#1\over#2}}         
\def\ninv{{1\over n}}
\def\sst{\scriptstyle}

\begin{document}

\bibliographystyle{plain}

\vskip 1pc
\nl

{\bf
\centerline
{SCALINGS AND PLASMA PROFILE PARAMETERISATION }
\centerline
{OF ASDEX HIGH DENSITY OHMIC DISCHARGES} }
\nl
\centerline
{P.J. Mc Carthy$^{1}$,
K.S. Riedel $^{2}$,
O.J.W.F. Kardaun,}
\centerline
{H. Murmann,
K. Lackner, the ASDEX Team}
\nl
\vskip 1 pc
\centerline
{Max-Planck-Institut f\"ur Plasmaphysik, EURATOM Association,}
\centerline{D-8046 Garching bei M\"unchen, Fed.\ Rep.\ Germany}
\vskip 1 pc
\leftline{$^1$On attachment from University College, Cork, Ireland}
\leftline{$^2$ New York University, 251 Mercer St., New York, USA}
\vskip 2 pc
\centerline{\bf Abstract}
\vskip 8pt
\ni


     A database of high density
($\sst .3 < \nb/10^{20}{\rm m^{-3}} < .8$),
low $q_{ a }$ ($\sst 1.9 < \qc < 3.4$),
Ohmic discharges from the ASDEX
experiment is analysed statistically \cite{KRML}. 
 Bulk parameter scalings and
parameterised temperature and density profile shapes are presented.
The total plasma kinetic energy, assuming $T_i=T_e$, scales as
 $\sst \nb^{\ .54\pm .01} {I_p}^{.90\pm .04 } $
 and is almost independent of $B_t$.
The electron temperature profile peaking factor scales as
$\sst {T_0^{3/2} /     <T^{3/2}>} = $ $ \sst
 .94(\pm.04){q_{ a }}^{1.07\pm.04}$
in close  agreement with the assumption of classical
resistive equilibrium.
In the inner half of the plasma, the inverse fall-off length for both
temperature and density has a strong dependence on $q_{ a }$, with
the temperature dependence being more pronounced.
Outside the half radius,
 the $\qc$ dependence  disappears but the density
profile broadens near the edge with increasing plasma current.
A second database of moderate density, moderate $q_{ a }$
discharges ($\sst .2 < \nb/10^{20}{\rm m^{-3}} < .4$,
$\sst  2.4 < \qc < 4.2$),
is presented for comparison.

\np
\section{\bf 1. INTRODUCTION}

     In most high magnetic field tokamaks, the experimental
Ohmic energy confinement time increases roughly linearly with plasma
density over a large range of densities.
In general, moderate field devices observe a similar
but weaker increase in confinement time at small to moderate
densities. Unfortunately, the Ohmic energy confinement time saturates
at higher values of the Murakami parameter
$  \nb R/B_t$~\cite{MCB}. 
 This phenomenon is termed density rollover
and, under usual operating conditions in ASDEX,
occurs at line-averaged densities around
$.25 \times 10^{20}  {\rm m^{-3}} $
and magnetic fields on the order of two Tesla.
(Pellet fuelled discharges and the recently discovered Improved Ohmic
Confinement regime on ASDEX \cite{MSJ} 
are not considered here.)
We present a
statistical analysis of the bulk parameter scalings and profile
shapes in this saturated high density regime.
For a detailed discussion
of the theoretical aspects of profile shape determination, the reader
is referred to \cite{KRML}, 
which we abbreviate as KRML.

In our datasets,
 neither of which extends into the linear regime, 
no significant change was observed in the scaling of the
total kinetic plasma energy
as the rollover density was crossed and exceeded.
Plasma energy and the Ohmic power scalings  nearly cancel, leaving   a
moderate $\qc$          dependence  ($\tau_E \sim \qc^{.3}$)
 in the confinement time.

To analyse the profile shapes, we fit
   all the measured profiles simultaneously by means of
 a radial spline function
each of whose     coefficients depends on the plasma parameters
$\qc,\ I_p$ and $\nb$.
This powerful technique enables us to quantify               the
                          various parametric dependencies as a function
of radius. For the electron temperature profile, we
     find that the profile shape variation
consists
almost exclusively of a $\qc$ dependence confined to  $r/a \le .60$.
The density profile also exhibits a $\qc$ dependence, though it is
weaker than that of the temperature.
Unlike the temperature, however, we find a broadening of the
density profile near the plasma edge with increasing plasma current.

Our research systematises and validates earlier
graphical and qualitative studies of ``profile consistency''.
(See \cite{ABE} 
for a review of this topic.)
The 
statistical approach enables us to make quantitative statements
about the relative strength of the interior $\qc$ dependence and the
weaker      exterior parametric    dependencies of the profile shape.

Previous authors have reported a variety of different results
on the peaking factor and the interior domain scalings. The
multiplicity of results arises from a combination of tokamak to
tokamak differences, different physics regimes,
systematic measurement errors, and sometimes also
 from    the use of  relatively unsophisticated analysis methods.
 We find, for instance, that normalising by $\sst <T^{3/2}>^{2/3}$
in the temperature profile peaking factor calculation
  removes the ambiguities associated with $\sst <T>$ and we obtain a
scaling closely consistent with the assumption of resistive equilibrium
between temperature and current density.
%
%
Using the result that the (inverse) fall-off length is solely
      dependent on $\qc$,                          we show
that the temperature profile shape  cannot be
Gaussian. 


In Sec. 2, we describe our    low $\qc$, high density and
moderate $\qc$, moderate density databases. In Sec. 3, bulk plasma
parameter regressions are presented, including new results for the
${\rm Wp}$ and $Z_{eff}$ scalings. In Sec. 4, peaking factors
and central temperature scalings are examined.
%
Sec. 5 summarises the spline model and
the statistical methods  used in the profile analysis.
Sec. 6 discusses a number of practical
techniques
to improve the            analysis.
One of these
is to separate the parameterisation of the profile shape and size.
%
%
Sec. 7 presents our experimental shape analysis results in detail.
A discussion and summary is made in Sec. 8.
In the appendix, we discuss the statistical significance of
each regression variable in a least squares regression model.
We note that Secs. 5 and 6 and the appendix may be
omitted by those readers who are interested only in the
results and not the statistical methods.

\section{\bf  DATABASE DESCRIPTION}

      The main (high density) database consists of 105 pairs of
       time-compressed  experimental
 electron temperature and density profiles from  50
 ASDEX deuterium discharges with a corresponding set of bulk plasma
 parameters and equilibrium flux surface information.
      Each compressed profile typically
 consists   of the time average of 12 consecutive `raw' profiles
sampled at 17 msec intervals using the ASDEX Nd:YAG (`YAG')
Thomson scattering
system \cite{RSM}.  
 All profiles are measured  during the current flat-top
Ohmic phase of the discharge. Apart from making the data analysis
more manageable, the preliminary compression had the effect
of strongly reducing  profile fluctuations due to sawteeth.
Such averaging is free from bias as long as the sawtooth period
$(\simeq 10 {\rm msec}) $ is incommensurate with the sampling period.
Some typical parameter values for this
database are the line-averaged electron density
 $\ \nb= .4  \times 10^{20}{\rm m^{-3}} $,
the plasma current $\ I_p = 400  {\rm kA}$,
and the cylindrical edge $ q_{ a } = 2.5 $.
 ASDEX discharges have a fixed circular geometry with
  $R_{plasma} = 1.65 {\rm m}$
and $a_{minor} = .40 {\rm m}$.
Table I(a) presents a more detailed summary of the database.
All datapoints consist of
                   double-null divertor discharges in deuterium
from three consecutive  shot days (404 to 406)
spanning a one week period in February
1986 during which titanium gettering was extensively used (about every
third discharge).
One of the experimental goals for these shot days  was a
density limit investigation, resulting in the
presence of a density ramp of
$.2 \ \ {\rm to } \ \ .4 \times 10^{20}{\rm m^{-3}s^{-1}}$
    for 10 out the 50 discharges, which
accounted for $\simeq 50\%$ of the datapoints.
For the affected discharges, this implies a typical density variation
about the mean  of $\pm 7\%$  for each   200 msec time window.
In selecting the discharges,
 care was taken to cover as wide a range in ASDEX control parameter
operating space as possible. The main toroidal field is relatively
seldom varied, however, and   two thirds 
of the datapoints are clustered at toroidal
 magnetic field values of $1.8{\rm T}$ or $2.2{\rm T}.$
  Nevertheless,
the distinction between $q_{ a } $, $B_t$ and $I_p$   scaling
will be quite apparent.
 Bolometric measurements showed   that radiation losses
  for these discharges
($10 \% <  P_{rad}/P_{Ohmic} < 40 \%,$  increasing with density) were
strongly localised at the plasma edge.

For comparison, we present a complementary analysis of a second
database (38 compressed datapoints from 38 deuterium discharges)
of moderate density ($ \nb \simeq .3 \times 10^{20} {\rm m^{-3}}$),
 moderate $q_{ a }$
($q_{ a } \simeq 3.1$), non-gettered, Ohmic discharges
which
is summarised in Table I(b). On  ASDEX, these densities
are            centred around the transition to the rollover regime.
These discharges were made over a six month period
spanning 64 shot days from August 1985
to February 1986 and were selected from  days where the vacuum vessel
was in a nominally normal state (no gettering, stainless steel walls).
At other times within this period, however,
 the experiment was run under a variety of
conditions including carbonised walls and three days of operating
with He as     working gas. In addition,
the vessel was open for several weeks in December 1985.
The YAG system was calibrated using Raman scattering from hydrogen
gas on two occasions during this period.
Our database contains no data quantifying the resulting changes
or residual effects of these operating periods on
the wall condition and, in particular, the condition of the
YAG diagnostic window itself.
 Hence we expect the residual error to
the fits to these experimental data to be larger. For these reasons,
we include the scalings of the
 second database       largely for purposes of comparison.
The distribution of datapoints for each
dataset is displayed in the combined Hugill plot (Fig. 1).

      The two databases differ in three parameters, $\nb$,
 $q_{ a }$, and wall condition (gettered /   non-gettered). Thus
we can view
the combined database as being   clustered in two cells of
    eight possible combinations. Indeed,      some scaling differences
between the two sets of data were  visually apparent in
preliminary efforts at fitting the combined dataset.
By analysing each database separately,
we are able to estimate secondary, weaker effects that
are not immediately apparent in a joint analysis. In
our case, these effects are weak current and density
dependencies of the outer section of the normalised density profiles.
If the two data sets were combined,
these weaker effects would be obscured by artificial
secondary scalings arising from the clustering mentioned above.

     As is typical of single machine databases,  geometric
parameters such as    the plasma position
and cross-sectional area vary
very little (about $ 1 \% $) and,
 once the channel positions have been mapped onto the normalised flux
surface coordinate ($0 \le r \le 1$),
 these variables are ignored in the analysis.
We assume that the macroscopic plasma state is essentially determined
by the line-averaged density $ \nb $
and two of the three parameters $I_p$, $\qc$ and $B_t$.
These parameters represent the major control parameters which the
experimentalist utilises to vary the plasma state in an Ohmic plasma.

The condition of the plasma wall may significantly influence
plasma performance. In an effort to include these effects, an
additional plasma variable such as
the effective ion charge $\Ze$, or the total Ohmic power $P_\Omega$
is sometimes used as an extra      independent variable.
Both $\Ze$ and $P_\Omega$
depend on the control parameters \conp,
and therefore are not purely
measures of the condition of the plasma wall.

{\it To examine the extent to which the Ohmic power and $\Ze$ vary
independently of the control parameters \conp,
we carry out a principal component analysis}
(PCA) \cite{MKB}: 
The correlation matrix
of the logarithms of
 $ \nb, \ I_p, \ q_{ a }, \ \Ze,$ and $ P_\Omega$ is
calculated and diagonalised (see Table II).
Each eigenvalue is interpreted as the sample
variance of the corresponding principal component over the database, and
 a small eigenvalue indicates a near collinearity between
the original variables. Since the principal components,
by construction, are statistically uncorrelated,
the sum of a subset of eigenvalues gives the cumulative
variance explained by the corresponding subset of principal components.
 We find that, for both databases, {\it over $97\% $
of the (standardised) data variation 
can be explained by the first three
components.}
The residual $3\%$ of the total variance is attributed  to noise in the
temperature measurements which affects $\Ze$ and noise in $U_{loop}$
which affects both $\Ze$ and $P_\Omega.$
Hence, we discard the                    principal components
having the two smallest eigenvalues.

The remaining eigenvectors span a three-dimensional  subspace.
This subspace can be efficiently represented by linear combinations of
any well-conditioned
set of three of the five variables.
That \conp constitute such  a
well-conditioned set
                             is checked            by re-doing
the PCA for these three alone.    The eigenvalues of the
$3 \times 3$
correlation matrix, also listed in Table II,
are all of the same order of magnitude, indicating that, unlike the
larger set, no near collinearity exists between them.


For scalings of the bulk parameters, we generally preferred to use
the logarithms of $ \nb $, $ B_t $, and $ I_p $
because the correlation between $ I_p $ and
$ B_t $ was much less than that of either variable with $ q_{ a } $
(see Table II(c)). However,
some scalings are given in terms of $\nb$, $\qc$ and $BI$, where the
latter parameter was chosen to maintain a set of
nearly uncorrelated bulk variables.
The dominance of $ q_{ a } $  and the weak $I_p$ influence
in determining
plasma profile shapes motivated us
to use  $\nb$, $I_p$ and $\qc$      for the profile shape analysis.

\section{\bf  BULK SCALINGS }

     In this section, we examine the scalings of the bulk plasma
variables. 
Let us briefly summarise our results: 
The scaling of the total
plasma energy as determined by kinetic measurements is virtually
the same in both databases, i.e. no transition
was observed 
going from the rollover to the saturated Ohmic
confinement (SOC) regime.
However, the diamagnetic and equilibrium mhd estimates of this parameter
show strong differences between the two databases.
The plasma energy and the
Ohmic heating power have roughly similar parametric dependencies,
which results in a  confinement time scaling consisting mainly of
 a rather                      moderate  $\qc$
dependence.
Power law scalings for
$Z_{eff}$ as well as for $Z_{eff}-1$ are
also 
presented.
The latter parameter scales like $1/\nb$, indicating that
the impurity density is independent of $\nb$.

We use the Spitzer value for
the effective ion charge  $ Z_{eff} $,
which is calculated  from the electron temperature profile assuming
resistive equilibrium (well satisfied for the current
flat-top profiles selected) and
Spitzer conductivity. We note that it cannot be used as an independent
parameter in the temperature profile regressions below.
Recent Bremstrahlung measurements on ASDEX \cite{SRR} 
show that for
the SOC                         regime (to which our high density
database belongs), $\Ze$ is very flat over most of the plasma radius,
though tending to rise strongly near the boundary. We adopt
the conventional assumption of zero radial dependence  here.
The ion density is calculated with the assumption that
the sole     impurity
is oxygen. The plasma kinetic energy,
${\rm Wp}_{kin} = {3 \over 2}\int{(n_eT_e + n_iT_i)dv}$,
is calculated by assuming the ion temperature is equal to the
electron temperature, $ T_i =T_e $. This assumption is justified
when the electron-ion energy exchange time is much shorter
than the energy confinement time.
 In the main         database,
the typical values are $\tau_{ei}= 5 \ {\rm msec}$
and $\tau_E = 75 \ {\rm msec}$.
In some of the moderate density
discharges, this condition is
not satisified. 

As an independent measure of the plasma energy content we also make
use of the diamagnetic flux measurement on ASDEX
whose interpretation is not affected by the uncertainty in $ T_i $
and is further simplified, in the case of Ohmic plasmas, by
the absence of pressure anisotropy.
The extreme sensitivity of the measurement
 $({{\psi_{dia}}\over
 {\psi_{tor}}} \simeq 10^{-4}) $
to such factors as slight mechanical
displacements of the diamagnetic loop       means,   however,
that  the typical error associated
with the derived value for beta poloidal is
$ \delta(\bpd) = \pm.05 $ . For $ \bpd \simeq 0.3 $ (the lower limit
in each database) this implies an error of some $\pm 15\% $ in the
diamagnetic energy and confinement time. For these databases, the
diamagnetic energy  is  systematically greater than the kinetic energy
 with an energy--independent offset $ \simeq6 \pm2 $ kJ (see Fig. 2).

A third measure of the plasma energy is derived from equilibrium
magnetic measurements, from which the parameter
 $\beta_{pol}+ l_i/2$ can be recovered.
To isolate $\beta_{pol}$, we need an estimate for $l_i/2$, which,
as is well known, cannot be determined
from equilibrium data in the case of circular plasmas.
From an earlier investigation, we use 
the following empirical relation from
a  parameterisation of current density profiles derived from
experimental temperature profiles and the assumption of Spitzer
resistive equilibrium: $l_i/2 = .332 + .199 \ln \qc $.

The regression models considered are of the form
$y = \sum\limits_i \alpha_ix_i + \epsilon$ where $y$ and $x$ denote the
logarithms of the dependent and independent bulk plasma variables
respectively. The root mean square error (RMSE) of the fit is
$ \sqrt{ \sum_{i=1}^N (y_i -  \yh_i )^2  /  (N-p) }$ where {\it p}
is the number of independent variables (including the intercept)
and $\yh_i$ is the fitted value of $y_i$.
As our response variables are natural logarithms,
the RMSE corresponds to a relative error in
the physical variable.
We also quote the squared multiple correlation coefficient  \Rs ,
which represents the fraction of total variance about the mean
accounted for by the fit: ${\rm R^2} = \sum\limits_i
 (\yh_i - \overline{y})^2 /  \sum\limits_i (y_i - \overline{y})^2 .$
In the bulk scaling results that follow, all physical variables are
expressed in the units of Table I.

{\bf Total Plasma Energy Scaling }

\noindent The plasma energy content from kinetic data for the
      main         database satisfies

  $  {\rm Wp}_{kin} = 133(\pm 6) \nb^{\ .54\pm .01} {I_p}^{.90\pm .04}
 {B_t}^{.06\pm .04} \quad {\rm RMSE}=.04 \quad {\rm R^2} = .965 $

\noindent   Similarly,  the scaling law based on the
secondary        database is

 $  {\rm Wp}_{kin} = 128(\pm 6) \nb^{\ .53\pm .03} {I_p}^{.80\pm .05}
 {B_t}^{.13\pm .05} \quad {\rm RMSE}=.04 \quad {\rm R^2} = .959$

\ni  The plasma energy content inferred from the diagmagnetic flux
 for the main         database is described by

 $  {\rm Wp}_{dia} = 130(\pm 6) \nb^{\ .40\pm .01} {I_p}^{.88\pm .04}
 {B_t}^{.13\pm .04} \quad {\rm RMSE}=.04 \quad {\rm R^2} = .943 $

\ni  whereas the secondary        database gave a  poorer fit:

 $  {\rm Wp}_{dia} =  83(\pm 10) \nb^{\ .21\pm .05}
{I_p}^{.77\pm .07}
 {B_t}^{.33\pm .08} \quad {\rm RMSE}=.06 \quad {\rm R^2} = .878$

\ni  The plasma energy content derived from equilibrium magnetic
 measurements for the main         database is described by

 $  {\rm Wp}_{\equil} = 156(\pm 5) \nb^{\ .41\pm .01} {I_p}^{.94\pm .03}
 {B_t}^{-.01\pm .03} \quad {\rm RMSE}=.03 \quad {\rm R^2} = .968 $

\ni  whereas the secondary        database gave

 $  {\rm Wp}_{\equil} = 255(\pm 28) \nb^{\ .26\pm .04}
{I_p}^{1.19\pm .06}
 {B_t}^{-.36\pm .07} \quad {\rm RMSE}=.05 \quad {\rm R^2} = .940$

 The criterion for `significance' of a regressor is
whether or not the regression coefficient is at least twice its standard
deviation.
In the appendix, the statistical background of this
simple rule and its relation with ${\rm R^2}$ 
is discussed.  

From the indicated uncertainties
(1 standard deviation) of the regression coefficients, 
we see, for instance, that the $B_t$ scaling for
${\rm Wp}_{kin}$
is insignificant for the main         database, while in the
 secondary  database   it is (just about) statistically 
significant.
In both cases, omission of $B_t$
affects the goodness of fit only slightly in absolute terms
(eq. (A7)   gives $\Delta({\rm R^2}) \simeq .001,\ .007$ respectively).

Because of the design of the
database, the correlations between
$n_e$, $I_p$ and $B_t$ are small (see Table II(c)).
This results in 
 small correlations
 between the regression estimates of the exponents
of $n_e$, $I_p$ and $B_t$.
 In  our regressions, all these correlations are
 less than 0.15 for the main and less than 0.35 for the secondary
 database. For assessing the significance of the
 difference between a postulated scaling law
 and our empirical scalings, such low
 correlations between the estimates can be neglected in practice.


The toroidal field shows the most extreme variation in the
six ${\rm Wp}$ scalings presented above.
There is also, however, a strong tendency for the $I_p$ coefficient
to move in the opposite direction to $B_t$ when going from one
scaling to another. This
suggests a representation with $q_a$. To preserve the near
independence of the regression estimates, we choose $BI$ as the
conjugate variable \cite{KTCNIN}.  
In the new representation, we have the following regression
coefficients for the six cases (same order as above)

$B_tI_p: (\quad .49,\ \ \quad.46,\ \  \  \quad.52,
\ \   \quad.55,\ \ \ \quad.48,\ \ \quad.42),$

$\quad q_a: (\ -.42,\quad-.32,\quad-.37,\quad-.20,\quad-.47,\quad-.75).$

\ni The errors in the coefficients
are roughly the same as in the $(I_p,B_t)$   scaling.
Obviously, to a good approximation,
the simple transformation
$\alpha_{BI}={1\over 2}
(\alpha_B+\alpha_I)$ and $\alpha_q = {1\over 2}(\alpha_B-\alpha_I)$
holds.
We roughly summarise:  ${\rm Wp} \simeq
     (BI)^{0.5\pm 0.05} q_a^{-0.5 \pm 0.25}$, where here 
the approximate {\it ranges} are indicated for the regression
coefficients in the six cases above.
Hence, at constant $\nb$, the
$BI$      scaling is nearly constant, but the
$q_a$ scaling varies considerably with the type of measurement
and with the database. 

The reasons for the observed $\qc$ scaling
differences may  be partly due to physics, and partly to systematic
errors. 
${\rm Wp}_{mhd}$ is particularly vulnerable to errors in the
$\simeq 50\%$ $l_i/2$  correction to $\beta_{pol}+l_i/2$.
 To get an     idea of the influence of this error,
        we performed a       sensitivity analysis. Using
$l_i/2 = .332 + \delta_1 + (.199+\delta_2)\ln q_a $,
we made a number of
${\rm Wp}_{mhd}$ regressions for different choices of $\delta_1$ and
$\delta_2$ in the range -.1 to +.1.
(Note that $\delta_1$ can describe an error in
 $\beta_{pol}+l_i/2$ as well as in $l_i/2.$)
 We found that the
exponent 
for $q_a$ in the ${\rm Wp}_{mhd}$ regression varied as
$-.47+ 1.3 \delta_1 - 1.6 \delta_2$ for the main database and
$-.75 + .43 \delta_1 - 2.0 \delta_2$ for the secondary database.
Thus an error of $-.1$ in the $\ln \qc $ coefficient for $l_i/2$
(we would expect this to be an extreme case)
gives $\qc$ exponents of $-.31$ and $-.55$ for main and secondary
databases respectively. We conclude that this error source
cannot reconcile the scaling of
       $\wmhd$ with $\qc$         for the secondary
database with the remaining five ${\rm Wp}$ scalings.

Now we turn to ${\rm Wp}_{dia}$. Suppose
 we have an offset $\delta_{\beta}$ in $\beta_{pol,dia}$.
 A sensitivity analysis showed that this roughly gives
an offset of $-2 \delta_{\beta}$ in the $q_a$ coefficient (for both
data bases). The density dependence is offset by
 $-\delta_{\beta}$ for   the
main data base and by $-0.5 \delta_{\beta}$ for the secondary database.
It is noted that the observed
$6 kJ$ systematic difference between the ${\rm Wp}_{kin}$ and
${\rm Wp}_{dia}$ measurements corresponds (at $I_p = .4$~MA)
to a difference in $\beta_{pol}$ of
$0.05,$  which  is  consistent with
the observed difference in $q_a$ dependence between ${\rm Wp}_{kin}$ and
${\rm Wp}_{dia}$ for either database. Similarly,
it explains half of the difference in $n_e$ dependence. It does not,
however, offer a satisfactory explanation for the $\wdia$ or $\wmhd$
scaling differences {\it between} the databases, for which we must
appeal      to the already referred to differences in operating
regimes    (gettered/non--gettered walls, high     densities/rollover
 densities, etc.). 

{\bf Volume Averaged Electron Temperature}

\ni  Both databases are in rough agreement with the
Pfeiffer-Waltz \cite{PW} 
and JET Ohmic scalings \cite{BBB}.  
The main         database satisfies

 $  <T_e> = 0.639(\pm 0.03) \nb^{\ -.52\pm .01} {I_p}^{.97\pm .04}
 {B_t}^{-.04\pm .04}
 \quad  {\rm RMSE}= .04  \quad {\rm R^2} = .947 $

\ni Similarly, the secondary        database satisfies

 $  <T_e> = 0.605(\pm 0.06) \nb^{\ -.61\pm .04} {I_p}^{.93\pm .06}
 {B_t}^{.04\pm .07}
     \quad  {\rm RMSE}= .05    \quad {\rm R^2} = .913 $


\ni Here, $B_t$ is insignificant for both databases and its
omission  has a very small ($\simeq .001$) effect on ${\rm R^2}.$
Though a natural candidate for determining the temperature, we have
not included $\Ze$ in the list of regressors, since, in our case,
it is derived, assuming  Spitzer resistivity, from the temperature
profile itself:
$$  Z_{eff} \propto {{\langle T^{3 \over 2} \rangle _{area}
V_{loop} } \over { I_p R_{plas}}}  
$$
\ni  (neglecting variations in the Coulomb logarithm).









{\bf Loop Voltage/Ohmic Power  }

\ni We present
the loop voltage scalings. The Ohmic power scalings differ
from these by exactly one power of $I_p$.
    For the main         database, the loop voltage scales as

 $  V_{loop} = 2.38(\pm 0.1) \nb^{\ .36\pm .02}{I_p}^{.12\pm .05}
 {B_t}^{-.27\pm .04}
     \quad  {\rm RMSE} = .04    \quad {\rm R^2} = .864 $



\ni Similarly, the secondary        database satisfies

 $  V_{loop} = 2.40(\pm 0.1) \nb^{\ .30\pm .02}{I_p}^{.19\pm .03}
 {B_t}^{-.31\pm .03}
    \quad {\rm RMSE} = .03     \quad {\rm R^2} = .896$

\ni
Due to the additional factor of $I_p$ on both sides,
we get necessarily higher \Rs \ values
(.935 and .984)
for the Ohmic power regressions
(The  RMSE values are unchanged).

{\bf Spitzer Z effective }

\ni For the main         database, the Spitzer $ \Ze $ scales as


 $  Z_{eff} = 2.90(\pm 0.18) \nb^{\ -.48\pm .02} {I_p}^{.67\pm .05}
 {B_t}^{-.34\pm .05}
  \quad    {\rm RMSE}= 0.05     \quad {\rm R^2} = .893$

\ni Regressing the impurity contribution to $\Ze$ gives  an entirely different scaling:

 $  Z_{eff} - 1 = 2.21(\pm 0.33) \nb^{\ -1.14\pm .04}
  {I_p}^{1.51\pm .12} {B_t}^{-.81\pm .12}
  \quad    {\rm RMSE} = .18 \quad  {\rm R^2} = .844 $

\ni For moderate densities, the Spitzer $ \Ze $ scales as

 $  Z_{eff} = 2.67(\pm 0.58) \nb^{\ -.69\pm .08} {I_p}^{.64\pm .11}
 {B_t}^{-.22 \pm .13}
 \quad {\rm RMSE} = .09  \quad {\rm R^2} = .714$

\ni The impurity scaling yielded

 $  Z_{eff} - 1 =  1.83(\pm 0.71)\nb^{\ -1.10\pm .14}
  {I_p}^{1.07\pm .19} {B_t}^{-.43\pm .21}
 \quad {\rm RMSE} = .16 \quad {\rm R^2} = .709 $


    Using the relation $n_{H;D}
  =  n_e - \sum_i n_iZ_i$ (i-summation over impurity
species only),
      $Z_{eff} \equiv \sum_jn_jZ^2_j/n_e$ (j-summation over all species)
      can be re-expressed as
$$  Z_{eff} = 1 + {{\sum_i n_i(Z^2_i - Z_i)} \over {n_e} }
$$
With this representation, we see that
{\it the $\nb$ exponents in the impurity scalings
                    suggest   
       an impurity density (almost) independent of the line density.}
  The                                    strong $I_p$ and $B_t$ scalings
 are not so readily interpretable.

{\bf Energy Confinement Time }

{\it The  $I_p$ and $\nb$ dependencies of the total
plasma energy and the Ohmic power approximately cancel to
leave a relatively weak $\qc$ dependence in $\tau_E$.}
This makes the energy confinement scalings less pronounced.
The main         database
 yielded the following   scaling for the
kinetic $\tau_E$ :

 $  \tau_{Ekin} = 56(\pm 5) \nb^{\ .18\pm .03} {I_p}^{-.21\pm .08}
 {B_t}^{.34\pm .08}
 \quad {\rm RMSE}= .07 \quad {\rm R^2} = .392$

\ni The secondary        database fit yielded

 $  \tau_{Ekin} = 53(\pm 6) \nb^{\ .23\pm .05} {I_p}^{-.39\pm .06}
 {B_t}^{.44\pm .07}
 \quad {\rm RMSE}= .05 \quad {\rm R^2} = .709$

As explained in the discussion of the ${\rm Wp}$ scaling,
we prefer the representation in terms of
$B_t I_p$ and $q_a$:

  $  \tau_{Ekin} = 68(\pm 4) \nb^{\ .18\pm .03} {(BI)}^{ .06\pm .06}
 {q_{ a }}^{.28\pm .05}
 \qquad {\rm (High \ \ density \ \ database)} $

 $  \tau_{Ekin} = 72(\pm 6) \nb^{\ .22\pm .05} {(BI)}^{ .02\pm .04}
 {q_{ a }}^{.41\pm .05}
 \qquad {\rm (Moderate \ \ density \ \ database)} $

\ni The confinement time derived from the diamagnetic measurement
of the energy content for the main         database scales as

 $  \tau_{Edia} = 69(\pm 3) \nb^{\ .04\pm .02} {(BI)}^{ .08\pm .05}
 {q_{ a }}^{.33\pm .04}
 \quad {\rm RMSE}= .06 \quad {\rm R^2} = .403$

\ni The secondary        database fit yielded

 $  \tau_{Edia} = 50(\pm 5) \nb^{\ -.10\pm .05} {(BI)  }^{ .11\pm .05}
{q_{ a }}^{.53\pm .06}
 \quad {\rm RMSE}= .06 \quad {\rm R^2} = .699$

\ni The confinement time
derived from equilibrium magnetic measurements
for the main         database scales as

 $  \tau_{\Eequil} = 77(\pm 3) \nb^{\ .04\pm .02}
  {(BI)}^{ .04\pm .04} {q_{ a }}^{.23\pm .04}
 \quad {\rm RMSE}= .06 \quad {\rm R^2} = .260$

\ni The secondary        database fit yielded

 $  \tau_{\Eequil} = 103(\pm 9) \nb^{\ -.04\pm .05}
{(BI)}^{-.03\pm .05} {q_{ a }}^{-.02\pm .06}
 \quad {\rm RMSE}= .06 \quad {\rm R^2} = .036$

%
The difference in $\qc$ scaling 
between the two
databases suggests the possibility of 
a quadratic (i.e. $\ln \qc \ln \qc $)
dependence \cite{KTCNIN}. 
No curvature in the  $\ln \qc $ dependence was apparent, however, in a
plot of  $\ln \tau_{Edia} $ for both databases with their respective
$\nb$ and $BI$ dependencies removed.
 Fig. 3 shows $\ln \tau_{Edia}/ (BI/.8)^{.08}(\nb/.4)^{\ .04}  $
for the main         database and
             $\ln \tau_{Edia}/ (BI/.8)^{.11}(\nb/.4)^{\ -.10}  $
for the secondary database.
 A similar picture
holds for $ \ln \tau_{Ekin} $, whereas
for $ \ln \tau_{Emhd} $ there appeared to be a deterioration at high
$q_a$ (figures not shown). 
Three points at low $\qc$ seem to be somewhat outlying in Fig. 3.
However, their removal does not change the $\ln \qc $ regression
coefficient by more than one standard deviation.
Unlike some previous
reported results
(see, e.g., \cite{MBD}  
section 4) we found
{\it no significant confinement time deterioration with density in the
SOC             regime}. Fig. 4 shows $\tau_{Ekin}$ (with $I_p$ and
$B_t$ dependencies removed) versus $\nb$ for both databases, i.e.
      $\ln \tau_{Ekin}/ (I_p/.4)^{-.21}(B_t/2)^{ .34}  $
for the main         database and
      $\ln \tau_{Ekin}/ (I_p/.4)^{-.39}(B_t/2)^{ .44}  $
for the secondary database.
Similar behaviour is observed for both $\tau_{Edia}$ and
$\tau_{Emhd}$.
For the difference in $q_a$ scaling, the reader is referred to the
sensitivity discussion in the subsection on total plasma energy.


  \nsect
{\bf 4. CENTRAL TEMPERATURE AND PEAKING FACTORS}

{\bf 4.1   Electron Temperature, Density and Pressure at r = 0.2}

     To enable reconstruction of
                 absolute profiles from the profile shape scalings
presented later, we regressed $T_e$, $n_e$ and $p_e = 1.5       n_e
      T_e$ at the $20 \%$ flux  radius. This normalisation  radius lies
inside the inversion radius for our $q_a$ range and also has data points
on either side of it.
(The YAG channel closest to the magnetic
axis lies typically on the $14 \%$ flux  radius.)
    The high density low $q_a$ database satisfies

            $ T_{e\ 20\%}=   .369(\pm .02)
 \nb^{\ -.49\pm .01}{I_p}^{\ \  .29\pm .04}{B_t}^{\ .71\pm .04}$
  $\quad {\rm RMSE}= .04 \quad {\rm R_2}=   .937$

           $  n_{e\ 20\%}=   .802(\pm .01)
 \nb^{\ \ \   .98\pm .01}{I_p}^{-.28\pm .02}{B_t}^{\ .21\pm .02}$
  $\quad {\rm RMSE}= .02 \quad {\rm R_2}=   .997$

           $  p_{e\ 20\%}=   .444(\pm .02)
 \nb^{\ \ \  .49\pm .02}{I_p}^{\ \  .01\pm .05}{B_t}^{\ .92\pm .05}$
  $\quad {\rm RMSE}= .05 \quad {\rm R_2}=   .925$

\ni while the secondary        database yielded

            $ T_{e\ 20\%}=   .404(\pm .04)
 \nb^{\ -.58\pm .05}{I_p}^{\ \  .31\pm .06}{B_t}^{\ .71\pm .07}$
  $\quad {\rm RMSE}= .05 \quad {\rm R_2}=   .879$

            $ n_{e\ 20\%}=   .922(\pm .05)
 \nb^{\ \    1.05\pm .02}{I_p}^{-.34\pm .03}{B_t}^{\ .13\pm .04}$
  $\quad {\rm RMSE}= .03 \quad {\rm R_2}=   .984$

           $  p_{e\ 20\%}=   .560(\pm .07)
 \nb^{\ \ \  .47\pm .06}{I_p}^{\ -.03\pm .08}{B_t}^{\ .84\pm .09}$
  $\quad {\rm RMSE}= .06 \quad {\rm R_2}=   .864$

\ni Most noteworthy here is the $\sqrt{\nb}B_t$ -- like dependence of
$p_{e\ 20\%}$ for both databases.
On regressing $p_e$ at each of the five most central YAG channels, which
typically lie between the $14\%$ and $25\%$ flux  radii,   we found
a similar absence of an $I_p$ dependence (coefficients ranged from
0. to .05 with a typical standard deviation of .05
for the main         database; -.13 to .18 with a
typical standard deviation of .075 for the secondary        database).
 The remaining 10 channels all exhibited strong
 $I_p$ dependencies (coefficients up to 2.0).      Hence,
we note that {\it the central electron pressure is independent of the
total plasma current.} 
Since the onset of $I_p$ dependence occurs for those channels whose
radii roughly correspond  to minimum values of the sawtooth inversion
radius ($r_{inv}(min) \simeq 1/\qc (max) = .30$ and .24 for the
main and secondary        databases respectively),
we speculate that
this $I_p$ independence is coupled to sawtooth stability and
sawtooth induced transport.

{\bf  4.2              Temperature Profile Peaking Factor}


Though we later analyse,  as a function of radius, the bulk parameter
dependencies of the local shape parameter $L_{T_e}^{-1}$,
                                         we present here results
for the usual single-parameter measure of the temperature profile
shape, namely the temperature profile peaking factor. We show that
 {\it the observed peaking factor is very close to that
expected assuming  a Spitzer resistive equilibrium}.

We first express the local cylindrical safety factor $q(r)$ as
$$q(r) = {rB_{tor} \over RB_{pol}(r)}
\ \ = {2\pi r^2B_{tor} \over \mu_0RI(r)} \ \ =
{2B_{tor} \over \mu_0R<J>_r}  \eqno(1)$$
where $<J>_r =
{1 \over \pi r^2} \int_{r'=0}^r J(r')2\pi r' {\rm d}r'$  is the
current density averaged up to radius $r$.

Assuming a classical resistive equilibrium,
we immediately have  that
 $${<T^{3/2}>_{r} \over <T^{3/2}>_{r=1} }
  \simeq  {q_{r=1} \over q_{r}} \eqno(2)$$
where the uncertainty arises from the fact that we neglect radial
variations (assumed weak) in $\Ze$ or the Coulomb logarithm.
 Since all profiles are sawtoothing, we have $\sst q(r < r_{q=1})
  \simeq q_0 $,  a constant with a value close to unity. Hence we
expect the $\sst 3/2$ moments peaking factors,
$\sst  {<T^{3/2}>_{r \simeq 0} /<T^{3/2}>_{r=1} } $
  to scale as $ q_a$.
Since the YAG channel closest to the magnetic axis lies typically on
the
$14\%$ flux radius,  $T_0$ is essentially an extrapolated quantity.
We present, therefore,   scalings for
  the LHS            of eq.         (2) at
both $r=0$ and $r=.2 $ where the latter radius satisfies the double
 requirement of being safely inside the inversion radius for all
 profiles as well as lying in the data region.
For the main         database, we obtained

$ {T_0^{3/2} \over <T^{3/2}>_{r=1}} = 1.02(\pm.06)\nb^{\ .03\pm.02}
 {I_p}^{.11\pm.09} {q_{ a }}^{1.13\pm.06}
 \quad  {\rm RMSE}= .06  \quad {\rm R^2} = .874 $

$ { <T^{3/2}>_{r=.2 } \over <T^{3/2}>_{r=1} } =
 .99(\pm.04)\nb^{\ .03\pm.015}
{I_p}^{.07\pm.06} {q_{ a }}^{1.07\pm.04}
 \quad  {\rm RMSE}= .04  \quad {\rm R^2} = .931 $

\ni while the secondary        database yielded

$ {T_0^{3/2} \over <T^{3/2}>_{r=1}} = .96(\pm.17)\nb^{\ .15\pm.11}
{I_p}^{-.12\pm.26} {q_{ a }}^{1.07\pm.14}
 \quad  {\rm RMSE}= .10  \quad {\rm R^2} = .795 $

$ {<T^{3/2}>_{r=.2} \over <T^{3/2}>_{r=1} }=
1.09(\pm.10)\nb^{\ .13\pm.07}
{I_p}^{-.07\pm.10} {q_{ a }}^{.95\pm.09}
 \quad  {\rm RMSE}= .06  \quad {\rm R^2} = .883 $

Since, for all four regressions, the $\nb$ and $I_p$ coefficients are
at best marginally significant at the $95\%$ level, the results indeed
strongly indicate that the peaking factor
   (eq. (2))   is determined solely
by $q_{ a }$.  We now assume a sole $q_{ a }$ dependence and re-do the
regressions to get a
tighter 
confidence interval for the constant
factor and the exponent. For the high density  database we get

$ {T_0^{3/2} \over <T^{3/2}>_{r=1}} =  .94(\pm.04){q_{ a }}^{1.07\pm.04}
 \quad  {\rm RMSE}= .06  \quad {\rm R^2} = .866 $

${<T^{3/2}>_{r=.2}\over <T^{3/2}>_{r=1}}
=.93(\pm.03) {q_{ a }}^{1.03\pm.03}
 \quad  {\rm RMSE}= .04  \quad {\rm R^2} = .925 $

\ni while the secondary        database yielded

$ {T_0^{3/2} \over <T^{3/2}>_{r=1}} = .85(\pm.11) {q_{ a }}^{1.13\pm.10}
 \quad  {\rm RMSE}= .10  \quad {\rm R^2} = .780 $

${<T^{3/2}>_{r=.2}\over <T^{3/2}>_{r=1}}
=.97(\pm.07){q_{ a }}^{.99\pm.07}
 \quad  {\rm RMSE}= .07  \quad {\rm R^2} = .870 $

In all four cases,
the $q_{ a }$ exponent is unity to within two standard deviations.
Similarly,  the constant factors  are within two standard deviations
        of unity (with the marginal exception of  
            the second regression: $.87 \le {\rm const} \le .99$).
These results are consistent with, but more specific than
the inequality
$$q_{ a }^{2/3} \le  {T(0) \over <T>} \le q_{ a }  \eqno( 3)$$
  derived
by Waltz et al \cite{WWG}  
using classical resistivity plus
sawtoothing.
Since we use        $\sst  <T^{3/2}>$
rather than $\sst <T>^{3/2}$,
the upper and lower bounds of the
inequality coincide in our case.
To enable comparison with previous work, we regressed
                                    the more traditional peaking factor
$\sst {T(0) /     <T>}$ to obtain,  
         for the combined database, the scaling:
$\sst {T(0) /     <T>} =$
 $\sst 1.00(\pm .028){q_{ a }}^{.73\pm .030}$
  whereas our normalisation gave (also for the combined database)
$\sst T(0)/<T^{3/2}>^{2/3} =$  $\sst .99(\pm .028)\qc^{.67\pm .029}.$
  {\it We believe      these results constitute
strong    evidence that the resistive equilibrium ansatz is
sufficient to explain quantitatively
the scaling of 
the  peaking factor
for ASDEX Ohmic temperature profiles.}

\nsect
{\bf  5. PROFILE SHAPE ANALYSIS}

The next two sections are devoted to a detailed description of
profile parameterisation techniques and the ensuing methods of data
analysis, tailored to our case. For a more theoretical background,
the reader is referred to KRML.
The experimental results are presented in section 7.

We assume that the logarithm of the temperature satisfies
$ \ln T(r,q_a,I_p,\nb)  = \mu(r,\xv) + \eps$ where  $\eps$ is a
random error. 
The deterministic part, $\mu(r,\xv)$, is represented as
a spline with reasonably high resolution in the radial direction,
and a simple (polynomial-type) dependence on the
plasma-parameters.
{\it The coefficients of the representation
are determined by fitting all profiles simultaneously
in a weighted least squares regression.}


The temperature and density measurements are obtained using the ASDEX
sixteen channel YAG Thomson scattering
diagnostic \cite{RSM} 
with a sampling rate of 60 Hz.
             This system consists of sixteen spatial channels
located in the vertical plane at ${\rm R}= 1.63 $ m.
They are spaced at approximately 4 cm
intervals from $Z= .200 $~m to $Z= -.394$~m.
                     We did not use the $16^{\rm th}$
channel which lies very close to or on  the separatrix, as the
measurement failed for the majority of  profiles in this database.
The  radius (averaged over all profiles) of the flux surface
 passing through each channel is
presented in column 2 of Tables III and V
 for the main and secondary        databases respectively.

{\bf  5.1. Radial Representations }

In this subsection, we discuss continuous radial representations
of     plasma profiles.
We consider data
consisting of $n$ separate  compressed profiles  of a spatially varying
 plasma variable such as temperature or density, at
$p$ distinct radial points. Each  compressed profile  is the average of
      $m = 12$ consecutive measurements taken at 17 msec intervals.
We do, however, make use of the uncompressed profiles for the purpose of
estimating the channel-by-channel raw measurement fluctuations
within each discharge. Thus our temperature data can be
described  by   $T_{i,j}(r'_l)$,
where $i = 1,.\ .\ .\ ,m$ labels the uncompressed timepoint,
 $j = 1,.\ .\ .\ ,n$ is the compressed profile index,
 and $l=1,.\ .\ .\ ,p$ denotes the radial channel number.
We make a preliminary transformation of the physical measurement
locations $r'_l$ to the
corresponding flux surface radii $r_l.$

Continuous representations have the following characteristics:
a) A large number of dependent variables, represented by point data,
 is replaced by a  
small number of coefficients which  nevertheless will be sufficient
to represent all significant features 
of the profiles. b) Profiles measured at
two different sets of radial locations may be compared.
 This is relevant, e.g.,
where we wish to compare YAG temperature measurements
with electron cyclotron emission (ECE) data measured at different
spatial locations.
c) Smoothness is imposed
in the belief that the profiles are in diffusive equilibrium.

Instead of fitting the profile itself, we choose to fit
its natural logarithm $Y$.
Minimising the residual of the logarithm of the plasma profile
corresponds to minimising the relative rather than the absolute error.
Preliminary comparisons with low order spline or polynomial
fits to the actual  profiles  revealed
that the logarithmic fit tended to have not only
smaller residual errors on the logarithmic scale but also
on the usual physical scale.
 This indicates that the  `exponentiated form' of the logarithmic
 fit is a better
approximation to the actual plasma profiles than a comparable low order
fit to the linear profile.
We note that the difference between logarithmic and linear fits
decreases as more regression
 parameters (either spline knots or higher order
polynomial terms) are added.
Logarithmic fits have several other advantages. Firstly, the predicted
profile can never be negative.
Secondly, well-known power--law type scalings reduce to linear models.
Finally, if the noise level is proportional to
the absolute value of the measurement
(an admittedly idealised situation), then, on the
logarithmic scale, unweighted least squares may be used.

      {\it
Spline representations, which we employ here for profile
 parameterisation,  give  flexibility in
choosing between local resolution and compact global representation.
 } 
The profile parameterisations presented in the present work
are based on 
twice continuously differentiable splines
with a selectable number of knots, $\nu.$ The profile 
is forced to be parabolic inside the first knot,
    i.e. the region enclosing the magnetic axis.  
  The radius is decomposed into $\nu +1$ regions with
knots at
$r_1, \ r_2, \ .\ .\ .\ ,\ r_{\nu}.$
Such a 
profile may be parameterised explicitly by:
$$ \mu(r) =
\cases{
\mu(r_1) + \mu''(0)(r^2 - r_1^2)/2&
 {\rm for} $0\le r \le r_1$  {\rm (Inner Region)}  \cr
\cr
\mu(r_1) + \mu''(0)(r^2 - r_1^2)/2 + c_1(r-r_1)^3&
{\rm for} $r_1  <  r \le  r_2$  {\rm (  Region 1)}  \cr
\cr
\mu(r_1) + \mu''(0)(r^2 - r_1^2)/2 + c_1(r-r_1)^3 \cr
\quad +c_2(r-r_2)^3& 
 {\rm for} $r_2  <  r \le r_3$  {\rm (Region 2)}  \cr
\cr
\quad \dotfill  & \dotfill                            \cr
\cr
\mu(r_1) + \mu''(0)(r^2 - r_1^2)/2 + c_1(r-r_1)^3 \cr
\quad  +c_2(r-r_2)^3 
+ . \ . \ . \ + c_{\nu }(r-r_{\nu})^3&
 {\rm for} $r_{\nu} <   r \le 1$  {\rm (Outer Region)}  \cr
} \eqno(4)$$

In more compact fashion, we express the spline $\mu(r)$ as
$$\mu(r)= \sum_{k=-1}^{\nu }
\alpha_{ k} H(r_k) \varphi_{k}(r) \eqno(5)$$
where $\nu$ is the number of knots used,
$\alpha_k$ represent the spline parameters:

$\mu(r_1), \mu''(0), c_1, c_2, \ . \ . \ . \ , c_{ \nu}\ ;$
$\quad \ \varphi_k(r)$ are the polynomials: $\varphi_{-1}(r)=1$,\
$\varphi_0(r)= (r^2-r_1^2)/2$,\ $\varphi_1(r)= (r-r_1)^3$,\
.. . . , \ $\varphi_{\nu   }(r)= (r-r_\nu)^3$, \ and
$$ H(r_k) =
\cases{
0 &
 for $ r < r_k \ $\cr
1 &
 for $ r \ge r_k \ $\cr
} \eqno(6)$$
is the Heaviside unit stepfunction
$(r_k = 0, 0, r_1, r_2, .\ .\ .\ r_{\nu})$ .

As well as the parabolic restriction near the axis,
a `natural' spline boundary condition, $\mu''(1)= 0$,
was imposed in practice.
The spline fits were carried out        using
the SAS REG procedure \cite{SAS} 
which contains
convenient possibilities for restricted regression.
A similar spline model for                transport analysis
      of individual profiles has been
 used in \cite{HSJ}. 

{\bf 5.2. Parametric Dependencies of Plasma Profiles}

Since the plasma profile shapes depend on the bulk plasma
variables such as $q_{ a }$, the spline coefficients will be functions
of these parameters. 
Since our database is not expected to contain sharp transitions in
behaviour in parameter space, a low order parametric representation
is expected to be adequate.
Therefore, {\it we approximate the smooth parametric
dependencies of the profile shape by linear  or possibly
quadratic polynomials in the logarithms of the bulk variables.}

Let ${\vec x} = (x_1, x_2, x_3)$
 be the vector consisting of the bulk plasma variables
 $ q_{ a } ,\  I_p , \ \nb $. 
We define the {\it linear basis functions $g_0(\vec x) = 1$ (intercept),
$g_j(\vec x) = \ln x_j/x^*_j $, $j = 1,2,3$; where $x^*_j$
is a representative value} of the variable $x_j$ in the database
of interest. For ease of comparison, we choose for both databases
the same normalising values
$q^*_a = 2.5$, 
$I^*_p = .4$ MA and $\nb^{\ *} = .4 \times 10^{20} {\rm m^{-3}}$,
although these do not constitute a typical parameter set
 for the secondary        database.
By normalising the bulk variables to $x^*_j$, the value of the
response variable  at the
intercept in the regression becomes the predicted value at
${\vec x} = {\vec x}^*$.

The full profile representation can be written as 


$$Y(r,\vec x)= \sum_{k,j} \alpha_{k,j}
 g_j(\vec x)H(r_k) \varphi_k(r) + \epsilon
\qquad \equiv  \mu(r,\vec x) + \epsilon               \eqno(7)$$
where $\epsilon$ is an error term, whose    structure will
be discussed in the next subsection.
The basis functions are as defined before.       Note  that
higher order terms can be included,
if necessary, by extending
the set of possible basis functions:
 $g_{j,l}(\vec x) = \ln x_j/x_j^* \ln x_l/x_l^* $, etc.

\ni
{\bf  5.3. Error Structures:  }

In this subsection, 
  we present and motivate the splitting-up of 
the total regression error into  contributions 
       that are attributable to different physical sources. An
efficient analysis
should take into account the particular features  of such an error
structure. In the next subsection, we discuss
the methods of estimation we applied in our case. 

We distinguish between several categories
of random profile variations.
We use the term
`internal' variations, to denote 
fluctuations on a time scale at least as fast as
that of the diagnostic sampling rate. These include statistical noise
from the measuring process and plasma fluctuations
 arising, in particular, from  the $m=1$ sawtooth instability.
Discharge-to-discharge variations are
changes in the plasma profiles not observed within a single
discharge. These discharge variations include effects such as
impurity accumulation on the diagnostic windows and
the condition of the plasma wall. These impurity and plasma
wall effects tend to vary to an even larger extent from
one experimental operating period to the next.
In addition, discharges separated by a
recalibration of the YAG system can exhibit
systematic differences in the measured profiles.

As discussed in \cite{MRKML}  
appendix B,
this hierarchy of temporal scales for plasma variation generates a
compound error structure which can be treated statistically.
We give a simplified discussion here.
For convenience, we assume normally distributed errors, although
this assumption can be relaxed in most of the discussion.
Since the  profiles in our database are already
averaged over twelve consecutive time samples (see introduction),
they no longer possess the
same variance as the original uncompressed observations.
For time averaged datapoints,
$\overline{Y}(r,\vec x ) = {1 \over m}\sum^m_{i=1}Y_i(r,\vec x )$,
the total unexplained variance
can be decomposed into:
%
$$\sigma^2_{tot} =
 {\sigma^2_{int} \over m} + \sigma^2_{dis} \eqno(8)$$

where $\sigma_{int}^2$ is the
variance           of the internal or
`within-discharge' 
fluctuations of the uncompressed profiles, $m$ is the number
of timepoints  in the compressed profile (in our case $m=12),$
and $\sigma^2_{dis}$ denotes the
variance due to discharge-to-discharge variations.

To estimate the within-discharge variance $ \hsig^2_{int}$,
we analysed the original uncompressed data and, for each channel,
calculated the empirical
variance for each 12-point set
separately. 
We estimated $ \hsig^2_{tot}$
by regressing the set of $n$ (compressed) datapoints for
each individual measurement channel against
the  bulk variables and noting the unexplained
variance. The difference, as given by
the third term in eq. (8), is an estimate of the
discharge-to-discharge variance. 


In Tables III(b) -- VI(b) columns 3 and 4,
                         one can see the estimates of
$\sigma_{int}$ (scaled for compressed profiles) and $\sigma_{tot}$
for each of the 15 channels.
It is clear from 
 the large channel to channel variation displayed in these
tables, that it would not      be   justified to make model assumptions
that $\sigma_{int}$ and/or $\sigma_{tot}$ are the same for all channels.
\ni
{\bf  5.4. Coefficient Estimation}

We determine the spline coefficients, including  parametric
dependencies, by fitting all profiles simultaneously
in a weighted least-squares regression where the weights, $W(r_l)$,
depend on the channel location.

We wish to determine that
vector of coefficients $\alu$  which minimises







$$ {\sum_{j,l}
\biggl ( Y_{j,obs}(r_l) - Y_{j,fit}(r_l,      \alu )\biggr )^2W(r_l)
 } \eqno(9)$$

\ni where $W(r_l),\ l=1,...,15$ are appropriately chosen weights
for each of the 15 YAG channels. We investigated
 two
approaches
for determining $W(r_l)$ . In the first method, we rely on
the total unexplained variance for each channel (as discussed
in the previous subsection) as a measure of the channel weighting:
$ W(r_l)=\hsig^{-2}_{tot}(r_l)$.
A second
approach to the selection of the regression weights,
is the iterative estimation
of the residual variance of the spline fit at each channel.
At the $k^{\rm th}$ iteration we have:
$$ {  \hsi^2_{(k+1)}(r_l)= \ninv {\sum_{j=1}^n
\biggl ( Y_{j,obs}(r_l) - Y_{j,fit(k)}(r_l,\hat{\alu}_k)
  -\delta \Yb_k(r_l) \biggr )^2  } } \eqno(10)$$
where $\delta \Yb(r_l)$ is a possible systematic bias in fitting the
$l^{\rm th}$ measurement channel which can be estimated by including an
indicator dummy variable for each of the 15 channels in the regression.
 The inclusion of this term
prevents undue downweighting of channels where
the parameterised profile may consistently fail
to match the observed data.  For the
first iteration, the variances $\sigs_{(1)}(r_l)$, are initialised
to unity (equal weights). The iteration is terminated after the
third iteration. We then have   $ W(r_l)=\hsi^{-2}_{(last\
 iter.)}(r_l)$. If the regression model and the assumed error
structure is correct, this is a likely to be a consistent and
efficient estimate (see e.g. \cite{P}).  
Nevertheless, we regard $\hsi^{-2}_{tot}(r_l)$
                    as a more robust estimate than $\hsi^{-2}_{(last\
iter.)}(r_l)$ since it depends
only on the fit by the bulk variables
whereas the latter estimate also depends on
the spline model and has the additional problem
of the strong (anti)correlation of outer channel residuals. Accordingly,
 we preferred       to use $W(r_l) = \hsig^{-2}_{tot}(r_l)$ .
        The dominant effect of this reweighting is to decrease
  the influence of the channels near the plasma
boundary where the relative
  error is  largest.

\ni
{\bf  5.5. Criteria for Additional Free Parameters}

In the context of profile parameterisation,
some relevant statistical tests
for the significance of including additional variables
are discussed in KRML.  
In the case of independent errors, these criteria are given by
the F test \cite{MKB} 
and Mallows $C_p$ statistic~\cite{DS}.  
Mallows $C_p$ statistic is the sum of the total bias in the
regression and the total variance of the predicted values. As more
free parameters are added, the bias decreases but the
variance increases. To determine whether to add another parameter,
one can
look at the change in the $C_p$ statistic or apply the F
test. 

It should be noted that in practice
these statistical approaches may
either under- or overestimate
the significance of the additional variables since the
correlations in the errors are neglected.
In addition, these tests
neglect systematic errors and assume that the ideal data,
without measurement errors, is exactly describable by the
regression equation under consideration. Thus many spurious
dependencies may be included and real dependencies missed
by unthoughtful or automated use of these methods.

\nsect
{\bf{ 6. PROFILE PARAMETERISATION TECHNIQUES} }

In this section, we discuss a number of practical aspects
   encountered during our investigation, which
     are expected to be useful in any 
 profile analysis.
Logarithmic representations are employed throughout,
for the reasons outlined in subsection 5.1.

{\bf 6.1. Polynomial Models}


Initial efforts
concentrated on fitting polynomial
representations of the form:
       $$T(r)= T_0 exp(ar^2+br^4+cr^6) $$ 
This   model  was successful in reproducing  the general properties
of the ASDEX profiles, but not detailed features. Sharp gradients
and local flattenings
(due, perhaps,  to magnetic islands) were poorly modelled.
A disadvantage of the above model, exacerbated by the addition of
higher order polynomial terms, is the insensitivity of the inner region
to $r^4$ and higher powers as well as the high degree of stiffness of
these polynomials. To enable us to fit each region of
the profile in moderate powers of the radial coordinate,
we turned to spline representations.

{\bf 6.2. Once and Twice Continuously Differentiable Cubic Splines}

Following \cite{KMLRG}  
we first used  a
         five parameter, two knot Hermitian spline, i.e. one with no
continuity requirement on the second derivative. This turned out to be
clearly       better in parameterising steep gradients and abrupt
spatial transitions in profile shape than the polynomial model.
By experimentation, we found that a total of four knots, requiring
seven regression parameters, gave a practical balance
between fitting accuracy and significance of the spline coefficients.


A serious disadvantage to the Hermitian spline emerged, however.
By allowing discontinuous second derivatives at the knots, continuous
transitions in plasma behaviour were modelled as sharp jumps
across knot boundaries.   This effect was especially prominent
in the slope of the inverse fall-off length as a function of $q_{ a }$.
After some investigation, we opted instead for the
twice continuously differentiable spline model, eq.~(4).

{\bf 6.3. Spline Knot Locations and Boundary Conditions}

The knot positions are chosen such that the measuring channels
are distributed roughly equally in the various regions
between and outside of the knot
positions. Too many knots result in spuriously oscillatory fits.
The knot locations were varied manually to achieve a near `optimal'
fit as determined by the balance between goodness of fit and
significance of the fit coefficients.
We  decided on  a set of five knots at
     the following locations:  $r_{knot} = .2,.3,.4,.5,.65.$

The innermost channel is typically located at $r=.14$ and the outermost
channel at $r=.89$ (Table III column 2). When third degree polynomials
were used in the innermost and outermost regions, the extrapolated
curves (to $r=0$ and $r=1$ respectively)
had unphysical oscillations. These oscillations were eliminated
by reducing the number of free parameters for these regions. Near
the origin, the profile was forced to be parabolic
(this is already enforced in eq. (4)).
The so-called
natural boundary condition, $\mu''(1)=0 $, was applied at $r=1$.
We investigated the regression
fits with the natural boundary conditions applied to (a)
all spline coefficients (b)     bulk parameter--dependent
spline coefficients only.
Case (a) resulted in considerably higher fitting errors
than case (b) for, in particular, the outer channels of the
 main     database  density profile fit.
However, we concluded that the improvement in case (b)  was at the
expense of  overfitting of the outer channels and we present
 here only the results of the fully applied boundary conditions.
  In a preliminary version of this work \cite{MRKML}, 
the results of case (b) are  presented. 
Thus our 5-knot set
      yields  a model of    seven 
    spline coefficients with one boundary constraint
           per bulk parameter basis function used in
    the fit.
The profile parameterisations presented later were carried out
using the linear basis functions $g_j(\vec x),\  j=0,1,2,3,$ only.
Some quadratic and cross terms were very significant in
preliminary regressions involving both databases simultaneously.
  For reasons given in the introduction, however,
      the results we present come from
 separate profile shape analyses for each  database.
For these regressions,
             second order terms were rarely significant and the
goodness of fit was scarcely affected by the restriction to
linear terms.
Using the three bulk parameters \conpr, we have a regression model
with       a total of
(intercept + 3 bulk parameters)
$\times$  (7 radial coefficients
- 1 boundary condition)
           $  = 24$
degrees of freedom to fit (e.g. for the main         database)
$15 \times 105 = 1575$
individual temperature (or density) data.
With the foregoing boundary conditions,
this spline representation tended to be 
     stable 
in extrapolating
profile behaviour into regions where there were no measurement channels.

{\bf 6.4. Normalisation }

The goal of our profile analysis is to determine
the dependence of the profile shapes, i.e. the functions
$L^{-1}_{T_e}(r) = {1\over T_e}{d\over dr}T_e$(r)  and
$L^{-1}_{n_e}(r) = {1\over n_e}{d\over dr}n_e$(r), $0 < r < 1$,
on the plasma parameters. 
This brings up the
problem of fitting the profile size.
We found that when we made a simultaneous fit
of the size                  and the shape
of the profile,
the residual sum of squares was dominated by the uncertainty in
fit of the profile size.
     Therefore, a model was fitted which provided a free parameter
     to fit each individual profile size.
This normalisation
procedure, justified because the
 profile size scales out of the profile shape definition above,
 causes very significant reductions in the residual sum of squares of
the shape regressions.

Originally, we normalised each profile by its line-average,
calculated from the spline fit. This 
greatly reduced,
but did not minimise, the residual error in
the profile parameterisation
since the line-average  is itself a function of the profile shape.
Instead,  we estimated the  profile       size parameters,
         using the SAS procedure GLM \cite{SAS},  
by treating the profile   index as an indicator
variable. This yielded as normalising factor
the radially independent term
in our spline repesentation, i.e. $\mu(r_1)$, the
profile value at the first knot which is sited at $r=.2$.
Normalisation has the effect of
reducing by one the number of degrees of freedom for
each individual spline. 
 Hence, the total number of degrees of freedom
for  the 
  5-knot spline with coefficients dependent on 3 bulk parameters
(as described in the previous subsection) is 
reduced from 
24 to 20. 

{\bf 6.5. Operating Period Indicator Variable}

In the course of determining $\s2trl$ for each YAG channel,
plots of residuals versus shot number revealed
that the secondary        database residuals, whose discharges spanned
a period  of over six months in contrast to the one week span of the
main         database, fell into four distinct groupings
which we ascribe to four distinct
experimental operating periods (see introduction).
This 4-cluster formation
          was observed for all 15 channels, although the pattern formed
by the clusters
differed for each channel. 
       To remove this operating period contribution to
the overall unexplained channel variances for this database,
 and hence to enable a comparison
to be made between the two databases, separate indicator
variables for each operating period were added in the individual
channel regressions used to determine $\hsig^2_{tot}.$
 These indicator variables were not, however, included in the profile
 shape          regressions. To do so would have required an
additional 60 independent variables (4 for each channel) which,
in our judgement, would have led to overfitting of the profiles.

{\bf 6.6. Examination of Outliers}

Apart from the operating period effects mentioned above,
 plots of raw versus fitted data for the same channel-by-channel
regressions revealed that
a small number of individual channel measurements from both databases
produced strongly outlying residuals (the worst case was one of
 8.5 standard deviations).
To arrive at a quantitative criterion for identifying
suspect data,
we analysed the Studentised residuals. A Studentised residual
 is the difference between the observed and fitted value, normalised to
the RMSE. For normally distributed errors,
they have  approximately  a standard normal distribution.
If we consider a single Studentised residual, the probability that it
lies outside $\pm c$ is $\epsilon,$ where $\epsilon \simeq 1 -
 (2\pi)^{-{1\over 2 }}\int^{+c}_{-c}e^{-{x^2/     2}}dx .$
    Considering now $n$ uncorrelated residuals together, we have that
the probability of all $n$ residuals lying {\it inside} $\pm c$ is
$(1-\epsilon)^n$. Hence the probability of at least one among $n$
Studentised residuals lying outside $\pm c$ is given by
$1-(1-\epsilon)^n = \beta$,  say.
 Provided the correct model is used to fit the data, we suspect  any
outlier  whose Studentised residual exceeds $\pm c_{\beta}$
for a suitably small $\beta$ (using $\epsilon = 1 - (1-\beta)^{1/     n}
\simeq \beta/n\ {\rm for}\ \beta<< 1,$  we invert the probability
integral to determine $c_{\beta}$).
We chose $\beta = 1\%$ which, for
$n = 105$ and $n = 38$ , yields $ c = 3.90$ and $c = 3.64$ for the
main and secondary        databases respectively.
 Using this criterion,
we identified 6 suspect outliers
from the main         database
 and 18 from the secondary,
amounting     to $0.2\%$ and $1.5\%$ of the data respectively.


Profiles containing any suspect observations were now
examined individually. In most cases it was visually obvious that
the affected channel was inconsistent with the rest of the profile,
and such observations were marked as bad data. 
One discharge accounted for the majority of the suspect data in the
secondary        database. On inspection, it was clear that
the quality of the profile data for this discharge was so poor, that
it was excluded entirely from the subsequent analysis, thereby reducing
the number of discharges from 38 to 37 for this database.
On the other hand, several suspect observations from a single profile
in the main         database
were not visually inconsistent with the rest of
that profile's data. On investigating further, it turned out that this
discharge had the highest $B_t$ value (2.73 T) in the
 database. This highlights the need to examine all suspect outliers
 individually, since the influential position of this
data  suggests an inadequacy in the model used to fit the data rather
than in the data itself, and its rejection would be quite unjustified.
The small number of           observations finally deemed
to be faulty (a total of 4 points affecting 3 profiles in the
main database; 12 points affecting 9 profiles in the secondary database)
were deleted from the regression.

{\bf 6.7. Measurement Asymmetries}

Ten of the fifteen YAG channels in ASDEX are located in
nearly symmetric positions with respect to the horizontal midplane.
By examining the residual errors for each channel separately,
an up-down asymmetry was found in the density profile
(up-down difference $\simeq 7\%$)
The asymmetry was nearly uniform on all five pairs of measurement
channels and was independent of plasma parameters. No significant
asymmetry was found in the temperature profiles.

A likely explanation for this asymmetry was found           after a
routine inspection this year (1990). A slight misalignment of the
laser beam direction against the entrance slits of the detectors
may   have decreased the sensitivity of the radial channels below
the equatorial plane.
Another possible cause is a spatially
non-uniform distribution of impurities on the diagnostic window.
Assuming that for the range of scattered laser light
detected by the system
($.8 \mu m \le \lambda_{scatt} \le 1.06 \mu m$),
the  impurity accumulation  causes a spectrally     uniform
reduction in transmission, then the          temperature,
which is calculated from the ratio of the scattering signals of
two spectral channels at the same spatial location,
is            insensitive to such asymmetries and only the density
is affected.

To estimate and correct for this density profile asymmetry, we expanded
the set of bulk parameter  dependent,
radial basis functions to include a single asymmetry indicator variable
 which takes the value of $+1$ for the channels above the
midplane and $-1$      below the midplane.
The final model, the results of which are presented in the
next section, accordingly increases its    degrees of freedom by one to
give a total of 21 for the density profile regressions (the temperature
model remains at 20 degrees of freedom).

\nsect
{\bf{  7. EXPERIMENTAL RESULTS} }

{\bf 7.1. Temperature Profile}

In this subsection, we present a detailed 
graphical representation 
of our results.
The dominant shape dependency is a   
peaking of the profile inside $r=.6$         with increasing $\qc$.
The profile shape shows very little dependence on any bulk parameter
 outside $r=.6 $. Thus our result is similar to
 Fredrickson's \cite{FMG}  
(profile shape invariant for $r \ge .4)$  and
Murmann's \cite{MW}  
(profile shape invariant for region outside influence of $q=1$ surface),
though our more elaborate model
enables us to examine more sensitively 
the extent of the $q_a$ dependent region.

In Figs. 5 and 6 we display, as a function of radius,
 the reference profiles and
           the parametric dependencies of the
  negative  inverse fall-off length (IFOL),
normalised to the inverse minor radius. Dashed curves indicate
                local $95\%$ confidence bands. The corresponding
global $95\%$ confidence bands are, in this case, approximately 1.5
times wider.   See KRML for
a 
 discussion on local and global confidence bands.
By the term
 reference profile we mean the evaluation of the parameterised
spline fit at the representative set of parameter values $\vec x = \vec
x^*$ as discussed in section 5.2. Since the vector of basis functions
$\vec g(\vec x) = (1,\ln x_1/x^*_1 ,\ln x_2/x^*_2 ,\ln x_3/x^*_3 )$
reduces in this case to $\vec g(\vec x^*)=(1,0,0,0)$, the reference
profile is just that described by the set of
 intercept, i.e. bulk parameter--independent, spline coefficients.

 The parametric   radial dependencies of the IFOL are obtained
by differentiating the parameterised spline representation for the
normalised profiles both radially  and with respect to the
basis function of the desired
parameter. Fig. 5(b), for example,  shows the radial behaviour of
the $q_{ a }$ dependence which, using eq. (7)
 (with $g_1(\vec x) = \ln q_a  - \ln q_a^*),$    
is given by
  $${\partial(-L_{T_e}^{-1}) \over \partial \ln      q_{ a } }
   = -{\partial^2 \mu (r,\vec x ) \over \partial g_1 \partial r}
   = -{\partial \over \partial r}
 \sum_{k} \alpha_{k,1      } H(r_k) \varphi_k(r) \eqno(11)$$
\noindent We see that at $r \simeq .35$, where the profile shape is
most sensitive to $q_{ a }$, a unit change in $\ln q_{ a } $ causes a
change of $\simeq  5$ in the negative IFOL
corresponding to
$12.5 \ {\rm m^{-1}}$ for an ASDEX minor radius of .4~m.
  Parametric dependencies of `experimental' point values
are also displayed. These are calculated by differencing the measurement
values of pairs of neighbouring  channels:
$L^{-1}_i  = { T_{i+1} - T_{i}  \over
{.5(T_{i+1} + T_i)(r_{i+1}-r_i)}} $ and regressing
these approximate IFOL's on \conpr.
 Such a point value has the
 advantage that it is more local than the continuous function
represented by eq. (11), but the disadvantage that
the signal to noise ratio will be lower. The mean value  will also
     be particularly affected by systematic errors in one or both of
the adjacent channels.

The negative IFOL profiles for
minimum,       reference, and maximum  $\qc$,
displayed in Fig. 5(a),    show that the temperature profile shape
for the main         database is
remarkably invariant outside $r=.6$.
This behaviour is broadly similar  for the profiles of
the secondary        database.
 The larger error bands of the latter, reflecting the substantially
higher regression error for this database, are due in large measure
to the already discussed problem of multiple operating periods for
this database.
The parametric dependencies of the IFOL profiles are detailed in the
remaining sub-figures.

The $\qc$ dependence of
the inverse fall-off length of both databases increases rapidly
and {\it reaches a maximum near $r=.35$.} The strength of this $\qc$
dependence then  decreases sharply,
and outside $r=.6$         the profile
shape is        independent of $\qc$. Between $.15 \le r \le .25$
the point estimates of  the
`experimental' IFOL             suggest a radially uniform
$\qc$ dependence. However, the large error bars allow for a slope
between roughly -10 and 10 on the scale of the plot.
Inside the first knot, our radial spline model consists of a
parabola with  $T_0'=0.$ This requires  the IFOL as well as each
individual parametric dependency
 to describe a straight line through the origin in this region.

The $I_p$ and $\nb$ dependencies are
much weaker.
Although, over some portions of the radius,
rejection at the $5\%$ level is marginally avoided,
      temperature profile invariance with respect to $I_p$ and $\nb$ is
generally seen to hold, within the experimental error limits, for all
radii. 
Figs. 7 and 8 show    reference  profiles and
     parametric radial dependencies for the normalised profiles
$\ln  \hat T_e(r) $  (where
$\hat T_e(r)=  T_e(r)/T_e(r=.2) $).  These are equivalent
            to the integrals of Figs. 5 and 6, with the integration
constant chosen to give $T_e(r=.2) = 1.$
The invariance of the profile shape outside $r=.6$          is
reflected here by the fact that all profiles are parallel in this
region (see Figs. 7(a), 8(a))
and all parametric dependencies (almost) horizontal
(see Figs. 7(b)-7(d), 8(b)-8(d)).

We now consider the hypothesis that     $T_e(r)$ has a Gaussian
        shape.
Since, to a high approximation, the profile shape has been shown to be
 solely a function of $\qc$, the most
general family of admissible Gaussian profiles      is:
$$T_e(r) =      T_{e,0 }(\nb,I_p,\qc){\rm exp}(-f(q_a){r^2 })
\eqno(12)$$
where $f(\qc)$ is a positive function.
 The negative IFOL                    would then satisfy
$$-L^{-1}_{T_e} = -{d \over {dr}}\ln T_e(r)  = 2r f(q_a) \eqno(13)$$
which describes a family of straight lines through the origin.
{\it It is clear from Fig. 5(a) (and Fig. 6(a))
that this hypothesis is false for our data.}
It is also visually obvious
   that even  the $\qc$ -independent part of the profile
($r > .6$) does not lie on       a  Gaussian. To quantify 
        this, we made a regression of channels
12 -- 15 (i.e. those roughly satisfying $r \ge .6$ ) normalised to
their line-averaged value versus a Gaussian function, which
                             gave $   T_e(r\ge .6)  = $ $3.76
   T^{    line\   av.    }_{e\ (r\ge .6)}
 exp(-2.53r^2)$ with a RMS relative
error of $5.4 \%.$ Adding a cubic term already
strongly reduces the error:
                                  $   T_e(r\ge .6)  = $ $1.88
   T^{    line\   av.    }_{e\ (r\ge .6)} exp( 1.48r^2 -3.62r^3)$
 with a RMS relative error of $3.8 \%$ (F value for cubic term $\simeq
400).$  Thus, even for the outlying portion of the profile, we can
categorically rule out a Gaussian shape.

 Figs. 9 and 10 show a sample experimental $T_e$ profile from each
database, representing extremes of the $\qc$ range covered by our data.
In each case, the predicted profile with $95\%$ global confidence bands
is also shown. Since we are concerned here with both profile size and
shape recovery, these predictions come from a parameterised spline
regression of {\it unnormalised} experimental profiles. Hence the
confidence bands are wider (by $\simeq 50\%$)  than those of the
corresponding normalised profile predictions.

 {\bf{  7.2. Temperature profile variance} }

We now examine how well the fitted profiles describe the data
both in terms of root mean squared error and    the
fraction of data variance explained by the model.  
We also give a break-down of the
 profile fluctuations into within-discharge and
discharge-to-discharge contributions.

Table III(a) presents some descriptive statistics for the temperature
profiles in the main         database
on a channel-by-channel basis.
Table V(a) displays the equivalent information
for the secondary        database.  The channels are numbered
according to their vertical position with channel 1 at $Z=.200$~m,
channel 6 at $Z=0$~m, and channel 15 at $Z= -.353$~m.
For each channel,
the mean normalised flux radius is presented, followed by the
mean temperature and the spread in keV.
 The spread
(this term
    is chosen to avoid possible confusion with  standard deviation  in
the sense of `regression error')
  is just the `standard deviation from the mean', i.e.
 $      spread  = \sqrt{ {1 \over (N-1)}
  \sum_{i=1}^N  (y_i - \bar y)^2 } $,
  where $\bar y$ is the sample mean.
Recall that before the regression, the profiles were normalised
by the size parameter from the SAS GLM procedure, i.e.
$T_e(r=.2)$, obtained from fitting each profile individually.
Columns 5 and 6 present the
mean and  spread  of these normalised profiles.
Channels 5 and 8 have very small spreads
since they lie closest to the normalisation radius.

As we are fitting on the
natural logarithmic scale, it is of particular interest
to tabulate the logarithmic, or relative, spread of the normalised
profiles as a  measure of the total variation of the data for each
channel.            This quantity, together with the channel-by-channel
noise level estimates and                        RMSE's resulting
   from    our spline parameterisation of the 
 temperature profiles
is presented   for the main         database
 in Table III(b) and for the secondary        database in Table V(b).
 In column 3, we display
$\hsig_{int}/\sqrt{12} \equiv \hsig_{int,\ comp.};$
the estimated standard deviation of the within-discharge noise
scaled  for    time-compressed profiles.
%
Column 4 tabulates
$\hsig_{tot},$
the total noise level (of the compressed profiles),
  estimated by   regressing $\ln T_e $   for each
channel on the bulk parameters.
The differences of the squares of the entries
in column 4 and column 3 are an estimate of the
discharge-to-discharge variance.



The ratio of the two noise estimates
lies in the range
$.2 \le \hsig_{int,\ comp.}^2/\hsig_{tot}^2 \le .5$
 indicating that the 
discharge-to-discharge variance is the dominant contribution to
           $\hsig_{tot}^2.$
Using eq.  (8) and the RMS values (over the 15 channels)
 for $\hsig_{int,\ comp.}$ and
$\hsig_{tot}$,  we find that for the main         database,
            $\hsig_{dis}(rms) \simeq     3.6\%.$
          For the secondary        database,
$.1 \le \hsig_{int,\ comp.}^2/\hsig_{tot}^2 \le .6$ and we have
$\hsig_{dis}(rms)\simeq 4.5\%$.

To indicate how much bias          is introduced by our spline model,
 we also carried out
channel-by-channel regressions of the normalised profiles.
           This is
 equivalent to using an interpolating radial  spline.
The RMSE's for these regressions appear in
Table III(b) column 5. The RMSE calculated for each channel from
    the parameterised spline regressions appear
 in column 6. The generally close agreement between columns 6 and 5
($\hsig_{spline         } (rms) = .030;$  $
\hsig_{channel                      } (rms) =.028)$
confirms the adequacy of the 5-knot spline model. The largest
discrepancies occur for channels 2 and 10 which form one of the
five          up-down symmetric channel pairs. On inspecting the bias
(systematic deviation from the regression line) for each  channel,
it was found that channels 2 and 10 had by far the largest bias
 $(-3.3\%$ and $+2.9\%$ respectively; the next largest channel bias was
$1.2\%)$,  indicating that at least one of these channels suffered from
 a systematic error of up to $6\%.$
In contrast to the main database, the spline regression errors
in  the secondary database
 (see Table V(b) column 6) are, in
general, much larger than the channel-by-channel errors  in column 5
($\hsig_{spline         } (rms) = .069;$    $
\hsig_{channel                      } (rms) =.039).$
This is explained by the fact that,
unlike the channel-by-channel regressions,
                    we did
not use operating period indicator variables in the  spline
parameterisation of the secondary        database profiles
(subsection 6.5).
For both databases, the regression errors generally increase for
outlying channels, reflecting a progressively deteriorating
signal-to-noise ratio.   This is
 due to  the decrease in scattered laser 
               light signal intensity
         with decreasing electron density.
  Note that channel 12 is an exception, with errors
similar to channel 15. This is consistent with the fact
that, whereas all other channels had three distinct spectral filters
 (normally offering the choice of the less noisy of
              two independent determinations of the
temperature),           channel 12, at the time the discharges for
our databases were made,   had only two.
{\bf{  7.3. Density} }

Since many of the results presented in the last section apply to
the density profiles as well, we only mention the differences.
Tables IV and VI contain the density statistics for
the main and secondary        databases respectively.

  Figs. 11 and 12 portray graphically our parameterisation of the
density profile local shape parameter  $-L^{-1}_{n_e} =
 -{\rm d} \ln n_e / {\rm d} r$
                                for each database. Figs. 13 and 14
show the integrals of Figs. 11 and 12, i.e. the normalised density
profile parameterisations.  Figs. 15 and 16 show sample experimental
$n_e$ profiles (with prediction profiles and confidence bands) for the
same discharges which provided the sample $T_e$ profiles shown in
  Figs. 9 and
10. Note the `jump' in the predicted profiles at $r=0,$  which arises
from the presence  in the regression
 of the density up-down asymmetry variable (subsection 6.7).

 The dominant  feature  of the density IFOL profiles
(Figs. 11 and 12) is  a $\qc$ dependence closely mirroring
that of the temperature IFOL, though at a reduced magnitude
$\bigl (-{\partial(L_T^{-1}) \over
 \partial \ln q_{ a } }(max) \simeq 5$;\ $-{\partial(L_n^{-1})
 \over \partial \ln q_{ a } }(max) \simeq 2\bigr )$.
 The radial region over which it  is significant
($.25 < r < .55$ for the main         database) is          smaller
                   than the equivalent region for the temperature.
Thus the variation of density profile shape with $\qc,$
 while not as dramatic as that
of the temperature, is  nonetheless considerable, as is evident in the
contrast   between     Figs. 15 and 16.
The magnitude of the density IFOL remains smaller than that of the
temperature over the entire profile. At $r=.9$ it  takes values
between 2 and 3 (fall-off lengths between 20 cm and 13.3 cm for
a minor radius of $  .4$~m)
 whereas the temperature fall-off length at $r=.9$ is nearly
fixed at $L^{-1} \simeq 5.5$ (see  Fig. 5(a)), i.e.
a           fall-off length of about $7$~cm for
$a= .4$~m.
As was the case with the temperature profile, the $I_p$ and $\nb$
    shape dependencies                    are weaker than that of $\qc$.
Near the edge, however, there is a statistically significant
{\it broadening of the density profile shape
with increasing current}.
(We note with caution, however, that
  the `experimental' IFOL datapoints suggest  that
 this current dependence is due solely to the outermost channel.)
In addition, some flattening of the density profile
with increasing $\nb$
occurs  
in the region  $.5 < r < .7$.

For both databases, as can
be seen from the RMS values for $\hsig_{int,\ comp.}$ and
$\hsig_{tot},$
the discharge-to-discharge variance (~$\hsig_{dis}(rms)\simeq 2.2\%$
 for the main database, and $\simeq 3.2\%$
for the secondary) forms the largest             contribution
to the total variance, as was the case for the temperature profiles.
          The overall spline regression RMS relative error
for each database (.032 and .065) is very similar to the
corresponding temperature value.

 A number of density profiles in both databases are slightly   hollow
in the region $.3 < r < .4.$  Since the
         set of reference profile parameters
($q^*_a = 2.5$, $I^*_p = .4$ MA
  and $\nb^{\ *} = .4 \times 10^{20} {\rm m^{-3}}$)
was not very typical for the secondary         database, this feature
looks somewhat     exaggerated for that database
                          (see Figs. 12(a) and 14(a)).

{\bf{  7.4. Electron Pressure} }

     The analysis of the (logarithmic) electron pressure profiles,
defined as $\ln P_e  = .4055 + \ln T_e  + \ln n_e $, offers additional
insight,  as
can be seen from the parametric dependencies shown in Figs. 17-20.
For the main         database,
the most striking feature is that the $I_p$ dependence is significant
over most of the radius. 
In the outer region of the plasma, there is 
a clear broadening of the pressure profile with the current, while
peaking occurs in the region  $.2 < r < .4.$
The  $\qc$ dependence of the inner  half of the profile
is very strong (temperature and density profile dependencies
reinforce each other) whereas the
$\nb$        dependence is little changed from that of the density
profile.



\nsect

\nsect
{\bf 8. DISCUSSION AND SUMMARY    }

\ni
{\it Bulk scalings} 

In the first part of this  work, we
presented and compared the scalings of  various global plasma parameters
for two complementary ohmic datasets. 
For the main, high density database, the volume-averaged temperature and
 three independent measurements
of the  total plasma         energy
                         depend on the plasma
current  $I_p$ and line-averaged density  $\nb,$
but, at constant $I_p$ and $\nb,$
 are practically unaffected by the toroidal magnetic field.
For the secondary database, $\wdia$ and $\wmhd$      show strong
$\qc$ scaling differences, while the $\wkin$ and $<T_e>$ scalings
       are almost the same as in the main              database.
The nearly linear current dependence for both the temperature
and the total energy is reminiscent of L mode scaling.

The Spitzer $\Ze$ depends on all three control variables. Regression
of $\Ze -1$ indicates, for both databases,
that the impurity density is almost  independent of $\nb,$  but
strongly dependent on both $I_p$ and $B_t.$
The electron temperature profile peaking factor
$\sst  {T_0^{3/2} /     <T^{3/2}>} $
 scales as
 $\sst  .94(\pm.04){q_{ a }}^{1.07\pm.04}$,
in close agreement with the prediction of classical resistive
equilibrium.

The strong dependencies on the plasma current $I_p$  of both the total
plasma energy ${\rm Wp}$ and the Ohmic power approximately cancel to
give a relatively weak current scaling
for  $\tau_E.$
Replacing ($B_t$,$I_p$) by ($\qc$,$BI  $) gives a $\tau_E$
scaling with a weak   dependence on $\nb,$
a moderate         dependence on $\qc$
and with no statistically significant $B_p I_t$   dependence.
In    the range
            $     .3 < \nb/10^{20}{\rm m^{-3}} < .8$,
we detected little dependence of the global confinement time on the
density. In particular, the
{\it decrease} in the rollover regime
with density as reported by \cite{MBD},
                                     was {\it not found}.
It should be remarked that the SOC data for $\nb > 0.5 \times 10^{20}
{\rm m^{-3}}$ in \cite{MBD},    first reported in \cite{WGB}, is
based on a single scan with $I_p= 0.42$MA, $B_t= 2.2$T. Obviously some
scans show      confinement deterioration, while others show a flat or
even weakly improving $\tau_E$ verus $\nb$ dependence.
    The same phenomenon can be observed in plots from the Doublet III
analysis \cite{EPR}. The reason for this variability
 in SOC density scaling
is at present unclear, and it requires further investigation.

\ni
{\it Profile analysis}  

A careful statistical analysis is necessary to determine the
radially varying parametric dependencies
of the profile shapes on the bulk plasma variables. By simultaneously
fitting all profiles with spline coefficients which depend on the
plasma variables, we have been able to examine profile dependencies on
a detailed,    quantitative level.
Based on this spline model, a
convenient graphical represention has been used to inspect
visually the influence of the various plasma parameters on the profile
shapes. 

An earlier study of ASDEX temperature profile
shapes \cite{MW} 
(for both Ohmic and neutral beam heated discharges) revealed
that the shape depends strongly on $\qc$ inside the sawtooth mixing
radius, but is almost independent of plasma parameters
outside `the influence of the $q=1$ surface'.
The results of
 our profile parameterisation  are roughly consistent with,  and
constitute a refinement of
   this analysis, for Ohmic profiles.

Except for a dependency of  the outer region of the density profile
shape on plasma current (and  to a lesser extent  on $\nb$),
the $I_p$ and $\nb$     dependencies of both the temperature and
density profile shapes are rather    weak in general. In most cases
the current and density dependencies are not significant  given
the error bars of the datasets.

In the interior, $\qc$ is the dominant       bulk plasma
parameter in determining
the temperature shape. By $r=.5$, however, this dependence has
weakened considerably and outside $r=.6$, as is clear from Fig.   5,
          the IFOL profile has an invariant shape.
%
We note that the extent of the $\qc$ -sensitive region is reasonably
consistent with the widest sawtooth inversion radius in each database
($r_{inv}(max) \simeq 1/\qc (min) = .54$ and .42 for the
main and secondary        databases respectively).

 Comparing Figs. 11 and 5, we see that the variation in the density
profile shape, while significant, is much weaker than that of the
temperature. This follows from the result that,
over the inner half of the radius,
the sensitivity of the density IFOL to $\qc$
 is only  $\simeq 40\%$ that of  the temperature.
The $\qc$ dependence is only significant for $.25 \le r  \le .5$.
In contrast to the temperature shape, which is unique outside
$r=.6$, the density profile broadens significantly near the edge
with increasing current.

The electron pressure IFOL exhibits a very strong $\qc$ dependence
in the inner half of the profile, while increasing $I_p$ causes
a   broadening of the outer half, a tendency which intensifies
approaching the plasma boundary.

Our findings are well described in terms of profile invariance
\cite{KRML, KMLRG} 
and in  quantitative
agreement with important   criteria for profile
consistency as described by \cite{ABE} 
  and developed by many authors.
However, we have not addressed the relative merits of profile
consistency versus local transport models \cite{BBB,CCC}  
containing    sawtooth effects. This issue could be addressed by
a statistical comparison of experimental profile dependencies with
the dependencies        predicted by local transport models.

\centerline{\bf ACKNOWLEDGEMENTS}

The work of PJM was performed under a EURATOM supported
reciprocal research agreement between IPP and University College, Cork.
The work of KSR was supported by the U.S. Department of Energy, Grant No
DE-FG02-86ER53223. We are indebted to the referees, who made several
critical
  suggestions from which the manuscript has profited considerably.

\nsect
\centerline{\bf APPENDIX  A}
{\bf ON STATISTICAL SIGNIFICANCE OF REGRESSION VARIABLES} 

  The significance of a    regressor  $x_j$ in the least squares model
can
be interpreted in terms of $ \hal_j/\hsi(\hal_j) $,
the ratio of the fitted coefficient to its standard error estimate.
 Under standard least squares assumptions, including
 (a) the correctness of the regression model and (b)  normally distributed
 errors in the dependent variable,
     the ratio  $ \hal_j/\hsi(\hal_j)$
has a Student's $t$ distribution
     under the null-hypothesis that  $ \alpha_j = 0 .$
For any statistic T  that has a $t$ distribution with $f$ degrees of
freedom, the following relation between the critical value ${\rm t}_{f,
\epsilon}$ and the `exceedence probability' or significance level
$\epsilon$ holds:
$$ P \biggl\lbrace \left| {\rm T} \right| > {\rm t}_{f,\epsilon}
          \biggr\rbrace   = \epsilon    \eqno(A1)$$

The null-hypothesis $\alpha_j =0$ is rejected, and the regressor is
considered significant, if
$        |\hal_j         /\hsi(\hal_j)| > {\rm t}_{f,\epsilon}$ for
some small value of $\epsilon$, say $5\%$.
For many degrees of freedom  ($f >   30$ usually suffices)
  as in the present case,  Student's $t$
can be well approximated by the normal distribution.
Thus we have ${\rm t}_{f,\ .05} \simeq 2.0$, and the significance
criterion is                $ |\hal/\hsi(\hal)|  > 2.0 $ .
 This result is no longer exact, though still approximate, for
 mild violations of  the normality assumption.        A seriously
 deficient model can, however, invalidate this interpretation of
 the coefficient standard errors.

It is often useful to have an estimate of the contribution to the
overall ${\rm R^2}$ from each independent variable.
Without loss of generality, we consider the multiple linear regression
problem
$$ y = \alpha_1x_1 + \alpha_2x_2
 +\ .\ .\ .\ + \alpha_px_p + \epsilon  \eqno(A2)$$
with centered dependent and independent variables.
If the independent variables are uncorrelated, i.e. if
               $<x_j,x_k>\  \equiv \sum_{i=1}^Nx_{i,j}x_{i,k} =
\|x_j\|^2\delta_{j,k},$ it is easily shown that the least squares
solution reduces to
$$\hal_j = {{<x_j,y>}\over{\| x_j\|^2~}}; \quad
\hsig^2(\hal_j) = {{\hsig^2}\over{\| x_j\|^2}}  \eqno(A3)$$

Here $\hsig(\hal_j)$
is the estimate of the standard error for the coefficient
estimate $\hal_j$ and
$\hsig^2 = \|y-\hat y\|^2/(N-p)$ is the mean square regression error.
From the definition of the $t$ statistic, we have that

$$ t_j \equiv {{\hal_j}\over {\hsig(\hal_j)}}
 = {{<x_j,y>}\over{\|x_j\|\hsig}}
= {\sqrt{N-p}\over{\|y-\hat y\|}}{{<x_j,y>}
\over {\|x_j\|}} \eqno(A4)$$

\noindent In geometrical terms,
 $<x_j,y>/\|x_j\| $ is
the projection of $y$ onto $x_j$ where $y$ and $x_j$ are vectors in
$\Re^N.$ Hence, adding $x_j$ to the regression model
 makes a fractional contribution to the total
variance of $$ \Delta({\rm R^2})_j =
{<x_j,y>^2 \over {\|x_j\|^2\|y\|^2 } }\eqno(A5)$$

\noindent Noting that         $$1- {\rm R^2} = {\|y-\hat y\|^2 \over
{\|y\|^2}} \eqno(A6)$$

\noindent we can use eqs. (A4) and (A6) to
     eliminate all terms involving $x_j$ or $y$ in eq. (A5)
and we finally obtain
 $$\Delta({\rm R^2})_j = {t^2_j \over N-p}(1-{\rm R^2})    \eqno( A7)$$
where the LHS denotes the decrease in ${\rm R^2}$ if the $j^{th}$
regressor is removed from the model.
 Note this relationship strictly holds only for uncorrelated regressors.
 If we now    sum     up all contributions, we obtain, using   Pythagoras' theorem,

$\sum_{j=1}^p <x_j,y>^2/\|x_j\|^2 = \|\hat y\|^2$ which leads to
the equality
$$ \sum_{j=1}^p t_j^2 = (N-p){{\|\hat y\|^2}\over{\|y-\hat y\|^2}}
= (N-p){{{\rm R^2}}\over{1-{\rm R^2}}} \eqno(A8)$$

\noindent
This formula is the analog of Weisberg's partition
of $C_p$ \cite{W1}. 
It provides a useful practical check on the applicability
of eq. (A7) when
 the regressors are correlated, which is  usually the case.

\nsect
\centerline{\bf APPENDIX B}
{\bf  COMPOUND ERROR STRUCTURES: TEMPORAL HIERARCHY}

To efficiently estimate the spline coefficients, $\alu$,
we try to model the actual covariance matrix for the errors.
The closer the assumed or
estimated $\Siuu$ is to the actual covariance error structure, the more
accurate the ensuing estmates for $\alu$ are.

The assumption of independent errors is not always justified.
In general, tokamaks possess a compound error structure.
The first level of errors are statistical fluctuations which vary
from time point to time point within a given discharge.
The next level consists of those errors which vary from
discharge-to-discharge
(we assume here that there is only one compressed datapoint per
discharge)
but remain constant within a given discharge. Finally,
there are variations which only change between operating periods
of a tokamak. We denote the covariance matrices of
for the radial fluctuations of each of these three types of errors by
$\Siuu^{int} , \Siuu^{disch} ,  \Siuu^{op} $ respectively.

We use a triple index, $(p,i,t)$ to denote a given profile timepoint
where $p$ indexes the operating period, $i$ the discharge
number, and $t$ the time.
Within a single profile timepoint, the individual radial measurements
are denoted by a fourth index, $l$. The cross-correlation of any two
pairs of profile measurements, $(p,i,t)$ and $(p',i',t')$ is given
by a $15 \times 15$ matrix, $\Siuu_{p,i,t,p',i',t'} $.

We assume that the errors do not depend on the plasma
parameters and that the covariance structure
does not vary between different blocks of data at each level.
The most general error structure of this form is

$${ \Siuu_{p,i,t,p',i',t'} =
\Siuu^{int} \delta_{p,p'} \delta_{i,i'} \delta_{t,t'} +
 \Siuu^{disch} \delta_{p,p'} \delta_{i,i'}
 + \Siuu^{op} \delta_{p,p'}  } \eqno(B1)$$

  We restrict our attention
to datasets consisting of a single operating phase. In this case
eqn. (B1) reduces to
${\Siuu^{int} \delta_{i,i'} \delta_{t,t'} +
 \Siuu^{disch} \delta_{i,i'}}$.  For simplicity, we assume
that each discharge consists of $n_t$ timepoints.

We estimate the within-discharge variance, $\Siuu^{int}$
empirically by calculating the time point average and
the time point variance for each discharge separately:

$${  \Sihat^{int}_{k,l} = {1 \over{n_d (n_t - 1)}}}
 {\sum_{i=1}^{n_d} \sum_{t=1}^{n_t}
(Y_{i,t}(r_k) - \Yb_{i}(r_k))
 (Y_{i,t}(r_l) - \Yb_{i}(r_l))
  } \eqno(B2)$$
where
$${ \Yb_{i}(r_k)  = {1 \over{n_t }}}
 {\sum_{t} Y_{i,t}(r_k)
  } \eqno(B3)$$

The disadvantage of analysing only the
time averaged profiles is that information about
the statistical fluctuations is lost. In an optimal
statistical analysis, all time points would be retained and
analysed simultaneously. This analysis may become unwieldy when
the number of timepoints is large. However, if the
timepoint variations are comparable to $n_t$ times the discharge
variation, a statistical
analysis based on structured covariance matrices is desirable.

Within a single operating period, the total variation between
datapoints is estimated by

$$ {
 \Sihat_{k,l} =
{1 \over n_d n_t-f}  {\sum_{i=1}^{n_d} \sum_{t=1}^{n_t}
 (Y_{i,t}(r_k) - Y_{fit}(r_k;\alu))
 (Y_{i,t}(r_l) - Y_{fit}(r_l;\alu))
  } } \eqno(B4)$$

where $f$ denotes the number of fitted parameters.
The fitted values, $ Y_{fit}(r_l;\alu) $ depend
on the values of the plasma parameters and therefore
implicitly on the indices, $i$ and $t$.
$ Y_{fit}(r_l;\alu) $ may be estimated either by
regressing each measurement channel separately or by fitting
all channels simultaneously using the spline representation.
The latter method will inflate the variance if the profiles
cannot be well approximated by the spline representation.

The discharge variance is computed by subtracting the
within-discharge variance as defined in eqn. (B2) from the
total datapoint variance defined in eqn. (B4).

In describing nested error structures of this form, statisticians
use the terms ``within discharge variation'' to refer to the
time point to timepoint variation and ``between
discharge variation'' for the discharge variation.

Several caveats must be placed on this procedure. First,
using too many or too few terms
in the regression analysis will artificially inflate the
variance estimates. Second, the errors in the estimates of the
variances tend to be rather larger unless a substantial number of
profiles are available.

 \nsect
\centerline{\bf APPENDIX C}

 {\bf{ ANALYSIS FOR RADIALLY CORRELATED ERRORS}}

When the random errors are correlated, the weighted least
squares estimator is consistent but not efficient, i.e.
as the number of datapoints approaches infinity,
the estimates of the regression coefficients converge
to their true value but the rate of convergence
is not optimal.

To increase the precision of the estimate, one tries to model the
actual covariance matrix for the errors.
The closer the assumed or
estimated $\Siuu$ is to the actual error structure, the more
accurate the ensuing estimates for $\alu$ are.
We continue to assume that the statistical fluctuations are
temporally uncorrelated and neglect the parametric dependencies.
However we now allow radial correlations in the fluctuations.
Since the dataset consists of $15n$ datapoints, the entire
covariance matrix is $15n \times 15n$. However we assume a block
diagonal form for $\Siuu$
of the form:

$$ \Si^{tot}_{i,k,i',l} = \Si_{k,l} \delta_{i,i'} \eqno(C1)$$

where $i,i'$ index the time-averaged profile and $k,l$ index
the channel number.

 This covariance matrix of the residual radial errors
may be estimated by:
$$
{  \Siuuhat_{k,l} =
{1 \over n}{\sum_{i}
 (Y_{i}(r_k) - \Yu_{i,fit}(r_k,\alu))
 (Y_{i}(r_l) - \Yu_{i,fit}(r_l,\alu))
  } } \eqno(C2) $$

In the previous sections, we have assumed $\Siuu$ is diagonal.
To examine whether our initial hypothesis of independent errors
is reasonable, we
perform a principal components analysis on $\Siuuhat$.

If the condition number (the square root of the ratio of
the largest to smallest eigenvalues)
or the ratio of the arithmetic mean to the geometric mean
of the eigenvalues
is approximately one,
then the previous analysis, based on radially uncorrelated
fluctuations, is justified.

If the condition number is substantially larger than one,
then the statistical estimating efficiency can usually be improved
by prescribing a functional form, $\Siuu(\thu)$ to model the observed
covariance. We can simultaneously estimate
<$\alu$ and $\thu$ using maximum liklihood estimates~[1].

\centerline{\bf APPENDIX D}
{\bf{  Does Radius of $\qc$-dependent
Region Contract as $\qc$ Increases? } }

Since the sawtooth inversion radius decreases with increasing $\qc$,
($r_{inv} \simeq 1/\qc$ \cite{14}), a natural hypothesis
 is that $r_{tran}$, the `transition radius' where the temperature
profile shape becomes independent of $\qc$ , also scales with
$1/ \qc$. This question cannot be decided on the evidence of
Figs. 2 and 3, beyond the assertion that for $r \ge .6$ the profile
shape is $\qc$~- independent for all values of $\qc$ in the two
databases. To make this analysis as radially localised as possible,
while still attempting to satisfy the conflicting requirement that
 individual channel noise be smoothed out, we discarded the spline
model used generally in this paper in favour of a moving  average -
type local quadratic fit (3 fit parameters) to successive groups
 of four neighbouring channels. To eliminate the
possibility that the higher density of channels for $r \le .5$ (due to
the 5 channels lying above the midplane) might have a bearing on
the results, we retained just the 10 channels on or below the midplane,
resulting in seven groups per profile.
By evaluating the local fits at intervals of $1 \over {40}$ radius
(only that radial interval lying between the
second and third               channel in each group was used),
 a set of smoothed IFOLs was accumulated for
 each of the 142 experimental profiles in the combined database.
These formed the working data for the following analysis.

To identify a possible $q_{ a }$ dependence of $\rt,$ we
assume that $\rt$ is some simple function of $\qc$ , e.g.
$\rt = \alpha_0 + \beta_0 \qc,$ where the shape of the temperature
profile is postulated to be $\qc$   dependent inside $\rt$ only. The
determination of the parameters $\alpha_0$ and $\beta_0$ is a nonlinear
problem which was handled by the following procedure:

(i) Choose candidates $\alpha$ and $\beta$ from a grid of possible values.

(ii) For each $(\alpha, \beta)$ fit all profiles simultaneously with the scheme
$$ L^{-1}(r,\qc) =
\cases{
  u(r,\qc)& {\rm for} $0\le r < \rt$  \cr
  v(r)& {\rm for} $r\ge \rt$  \cr
} \eqno(20)$$
 where $\rt = \alpha + \beta\qc$  varies with each individual value
 of $\qc$ , and $u(r,\qc)$ and $v(r)$ are quadratic functions of $r$:
  $u(r,\qc)= a_0 + a_1\qc + (b_0 + b_1\qc)r + (c_0 + c_1\qc){\rm r^2}$
and $v(r)= d_0 + e_0r + f_0r^2$
with  zeroth order continuity  imposed at $r=\rt$ .

(iii) Store $\alpha$, $\beta$, and the mean square error (MSE) for each fit.

(iv) Locate
${\hat \alpha_0}$ and ${\hat \beta_0}$
 giving the best fit (i.e. the global minimum MSE value).
 
(v) Make a contour plot of  MSE
as a function of $(\alpha, \beta)$ and hence determine the $95\%$
confidence contour for ${\hat \alpha}$ and ${\hat \beta}$

(vi) If $ \beta = 0 $ lies inside the $95\%$ confidence
contour, then the null-hypothesis $\beta_0 = 0$ is compatible
with the data.

\noindent The contour plot obtained using the above procedure is
shown in Fig. 14. The global minimum MSE, $\hsig^2_{min}
= .0906$ , is given by $\rt = .620 - .012\qc$ .

To determine the $95\%$ confidence contour for $\beta$  where
 $\alpha$ is arbitrary, we use the result that the
relative difference in
the sum of squares has asymptotically a $\chi^2$ distribution with, in
our case, one degree of freedom (see \cite{5} Chap. 5).
For this analysis, because of the  restrictions used in
generating the input IFOL values, we have effectively 9 independent
measurements per profile, giving a total (for 142 profiles) of
$n=1278$.
From \cite{5} Table C.1 we have ${\rm \chi^2_1(95\%)= 3.84}.$
This gives $\Delta({\rm RSS})_{95\% {\rm contour}} =
 3.84 \times \hsig^2_{min} = .348 $ .
Hence the $95\%$ contour has the values $\hsig^2_{95\%} =
 .0906 + .348/1278= .0909$ .  This  MSE value is enclosed
  by the $3^{\rm rd}$ contour in Fig. 14 which leads to a
(slightly conservative) $95\%$ confidence band for $ \beta_0$ of
$(-.016,-.006)$. For                $\rt = {\hat \alpha_0}
 + {\hat \beta_0}\qc$ we have the extreme values $\rt
(\qc=1.9) = .60$ and $\rt(\qc=4.2) = .57$ . A search over the entire
 region bounded by the $95\%$ confidence contour yielded
$\rt(\qc=1.9)_{max} = .61$ and $\rt(\qc=4.2)_{min}= .57$.
Thus, although the null-hypothesis $\beta_0 =0$
was not satisfied, we have established < a $95\%$ confidence interval $(.57, .61)$ for $\rt$ for
the combined database where $\qc$ varies from 1.9 to 4.2. This small
variation in $\rt$ ($r_{trans,max}/r_{trans,min} \simeq 1.1$)
clearly contadicts the hypothesis that $\rt \propto 1/\qc$ since we
would expect, for our database, that $\rt$ would vary by a factor
of ${4.2\over {1.9}} = 2.2.$ We can offer no convincing explanation
for this result.



\section*{Supplementary material}

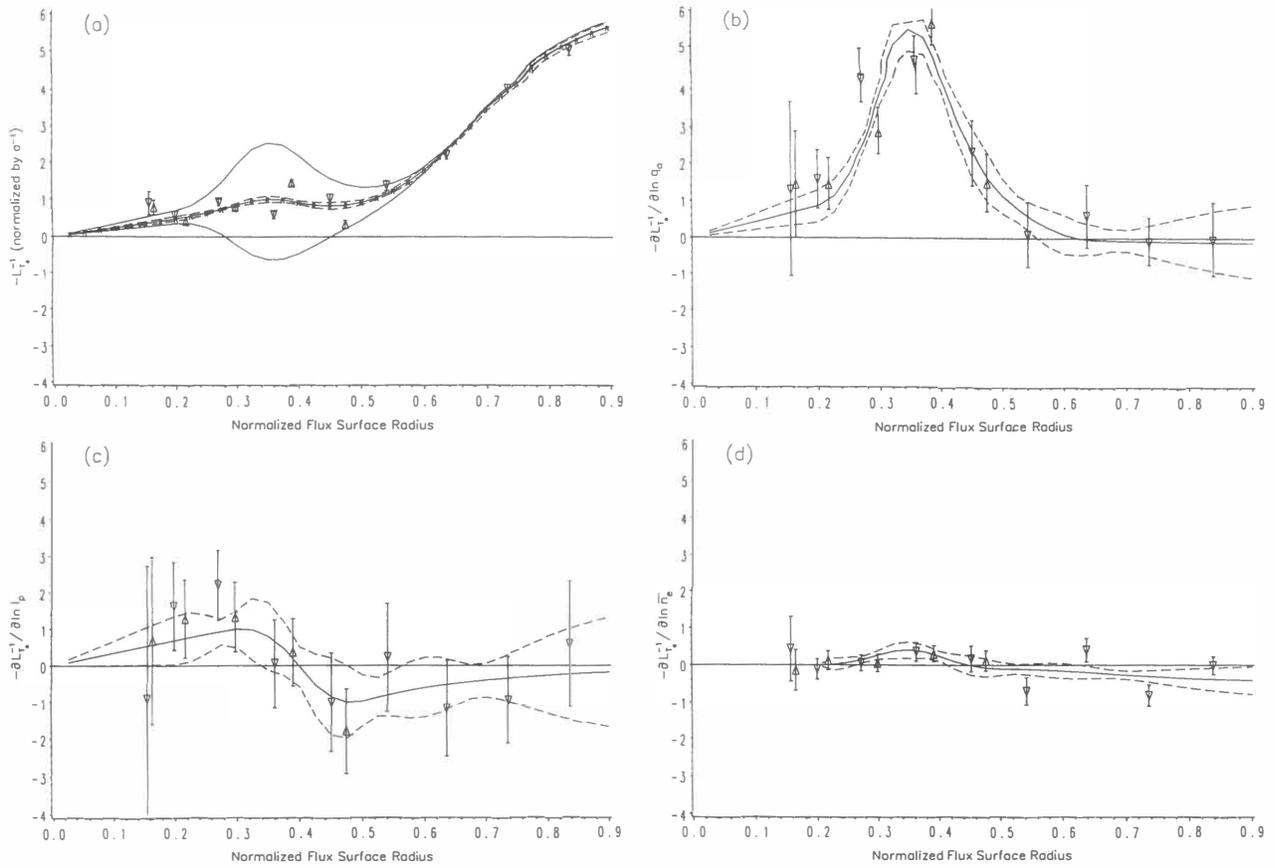

FIG. 5. *Reference negative IFOL profiles for the main (high density) database $T_e$, and parametric dependence profiles for $q_a$, $I_p$ and $\bar{n}_e$ with local 95% confidence bands. Reference values for discrete (two-channel) IFOLs and the corresponding parametric dependences are also plotted as individual data points. The central solid curve in (a) is the predicted negative IFOL at the reference parameter values $q_a = 2.5$, $I_p = 0.4$ MA, $\bar{n}_e = 0.4 \times 10^{20}$ $m^{-3}$. The local 95% confidence bands are plotted as dashed curves. The flattest and steepest predicted negative IFOL profiles in the database (i.e. those for minimum ($q_a = 1.9$) and maximum ($q_a = 3.4$) values of $q_a$) are plotted as solid curves without confidence bands. The individual data points are predictions for discretely calculated IFOLs (Section 7.1), again evaluated at the reference parameter values and accompanied by 95% confidence bands. (b) Parametric dependence of $L_{T_e}^{-1}$ on $q_a$. The solid curve is the radial profile of the change (in units of 1/minor radius) in the negative IFOL per unit change in $\ln q_a$ (see Eq. (11)). The individual data points give similar information on the discrete IFOLs. (c) Parametric dependence of $L_{T_e}^{-1}$ on $I_p$ and $\bar{n}_e$. (d) Parametric dependence of $L_{T_e}^{-1}$ on $\bar{n}_e$.*

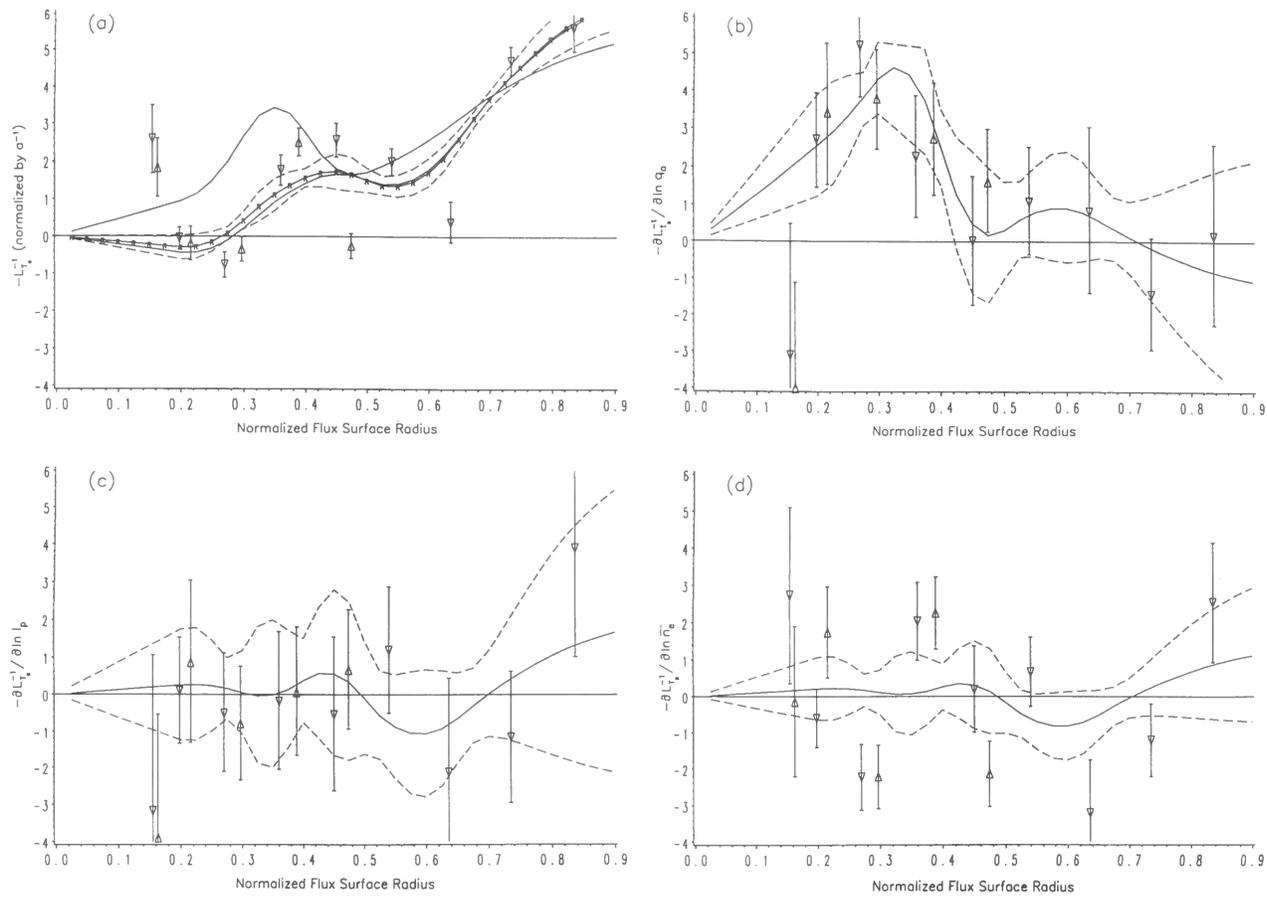

FIG. 6. *Reference negative IFOL profiles for the secondary (moderate density) database $T_r$, and parametric dependence profiles for $q_a$, $I_p$ and $\bar{n}_r$.*

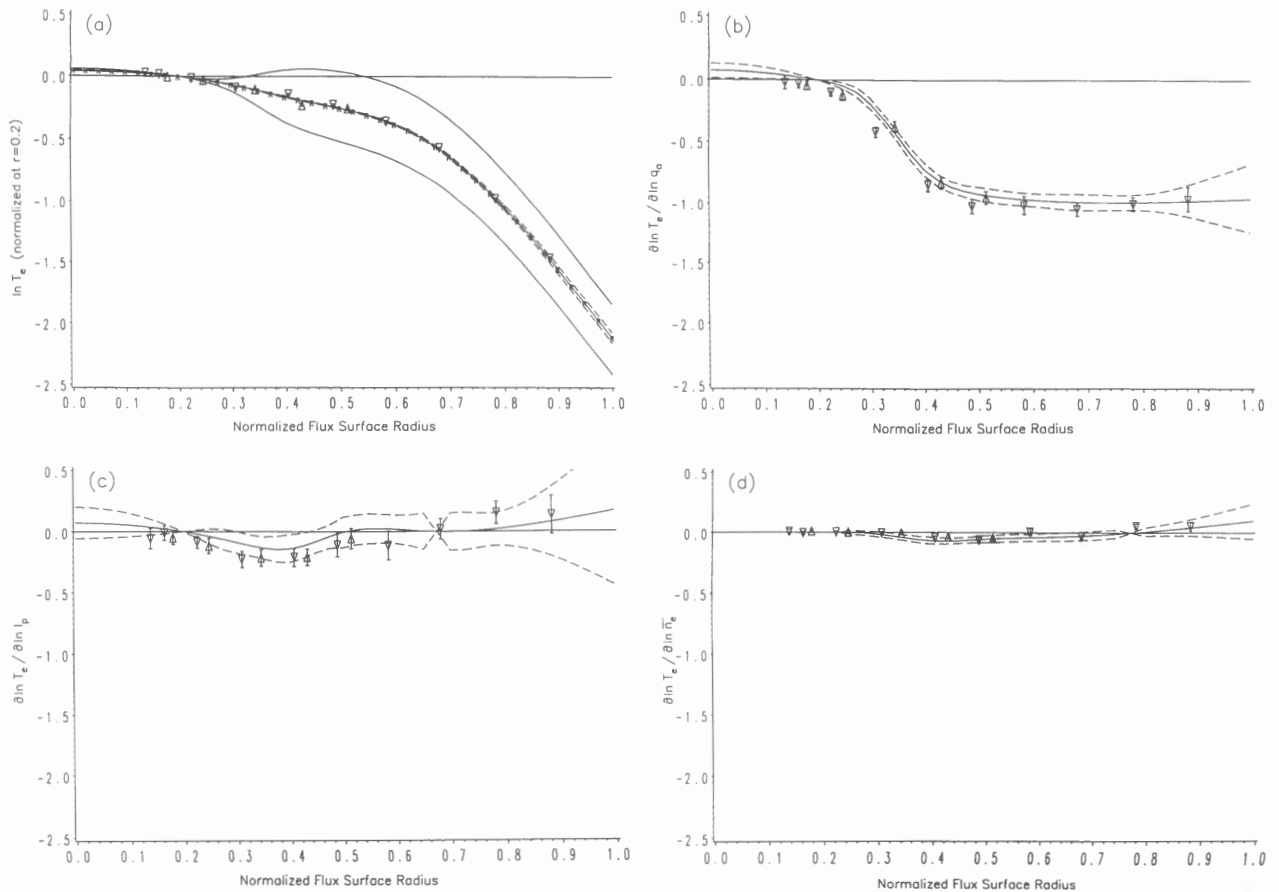

FIG. 7. Reference $\ln T_e$ profiles (normalized at $r = 0.2$) for the main database, and parametric dependence profiles for $q_a$, $I_p$ and $\bar{n}_e$.

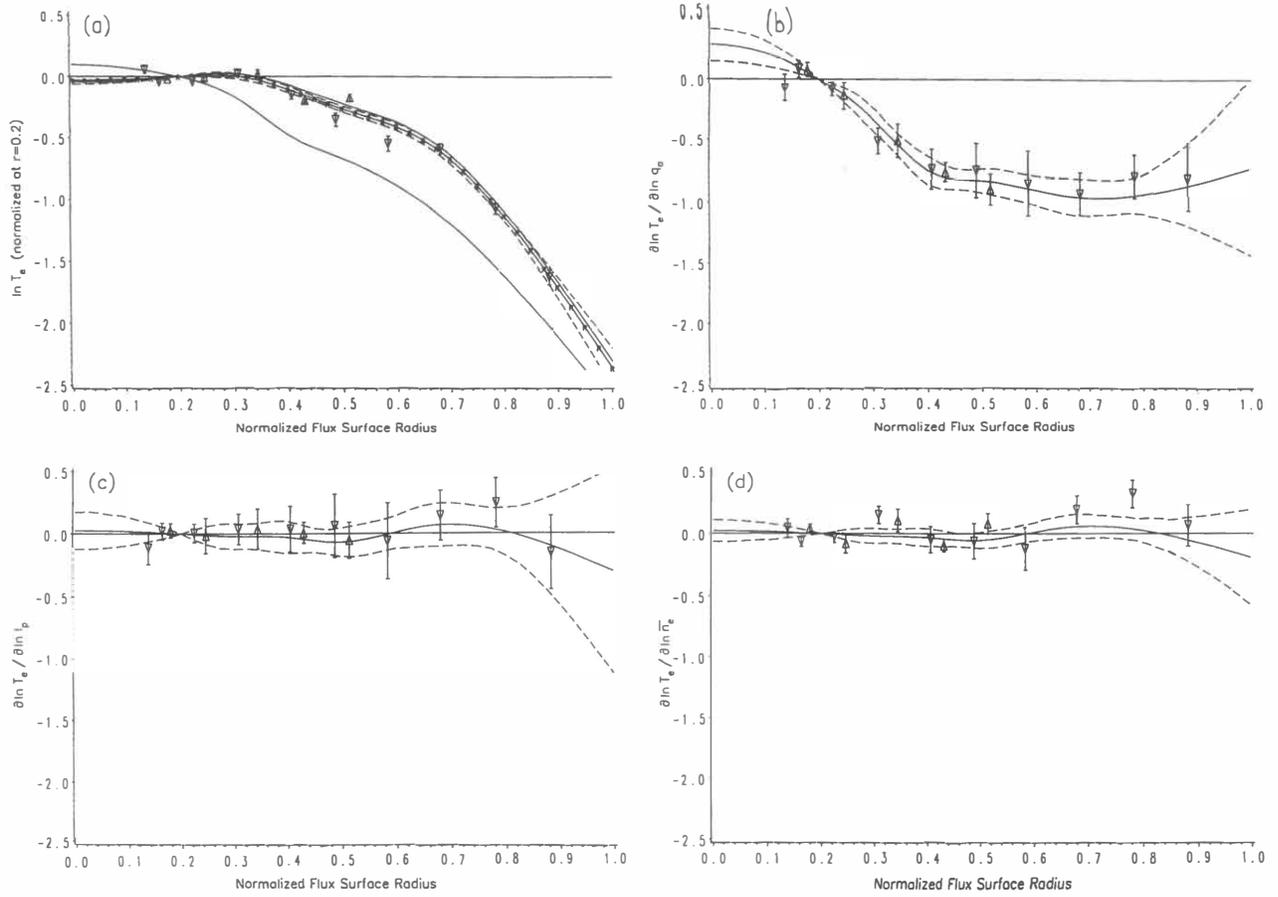

FIG. 8. Reference ln $T_e$ profiles (normalized at r = 0.2) for the secondary database, and parametric dependence profiles for $q_a$, $I_p$ and $\bar{n}_e$.

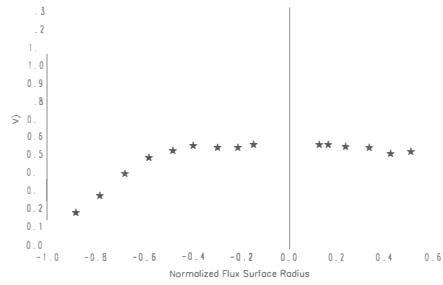

FIG. 9. *Sample experimental and predicted temperature profile with global 95% confidence bands from the main database. Parameters: $q_a = 1.94$, $I_p = 0.452$ MA, $\bar{n}_e = 0.704 \times 10^{20}$ m$^{-3}$.*

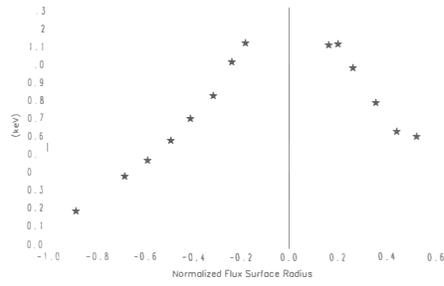

FIG. 10. *Sample experimental and predicted temperature profile with global 95% confidence bands from the secondary database. Parameters: $q_a = 4.01$, $I_p = 0.281$ MA, $\bar{n}_e = 0.299 \times 10^{20}$ m$^{-3}$.*

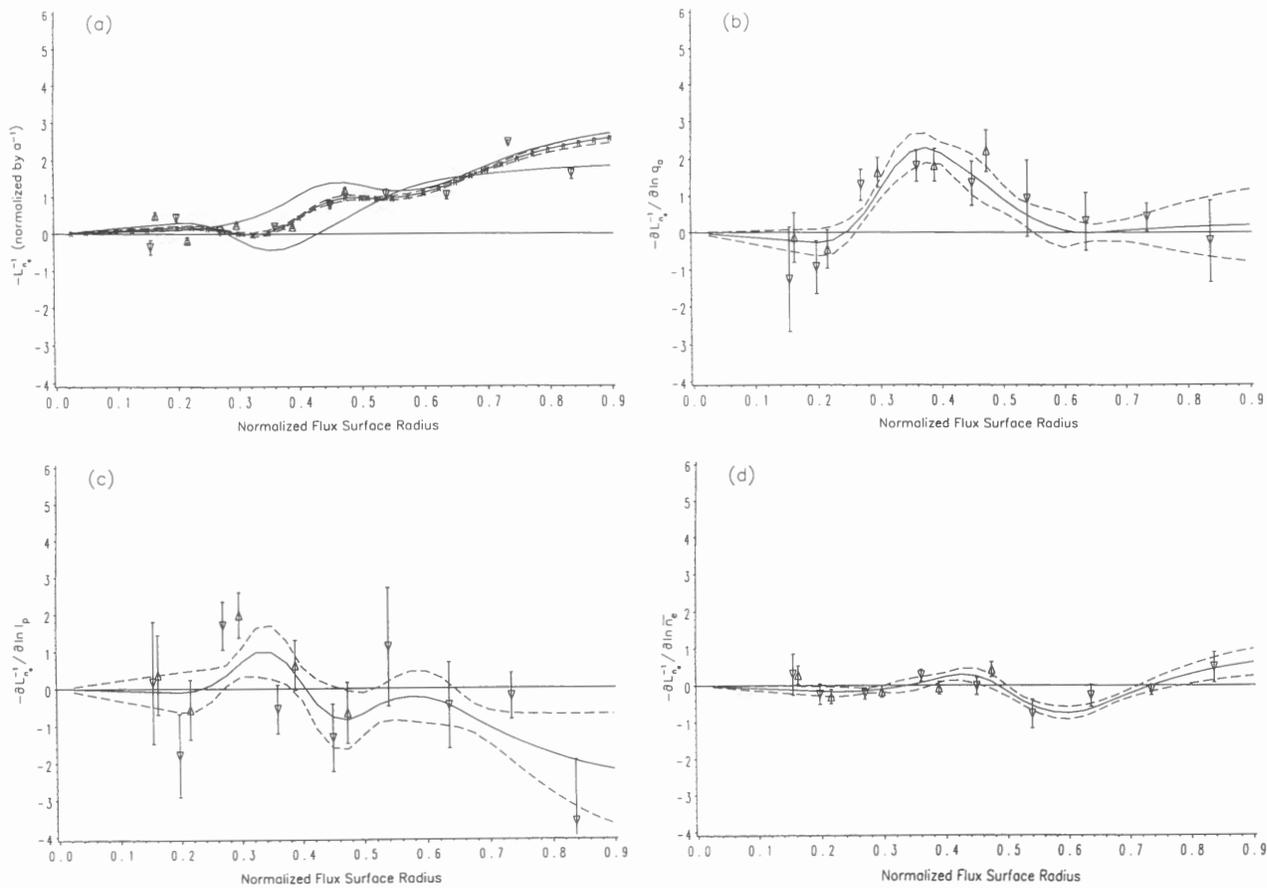

FIG. 11. Reference negative IFOL profiles for the main database $n_r$, and parametric dependence profiles for $q_a$, $I_p$ and $\bar{n}_r$.

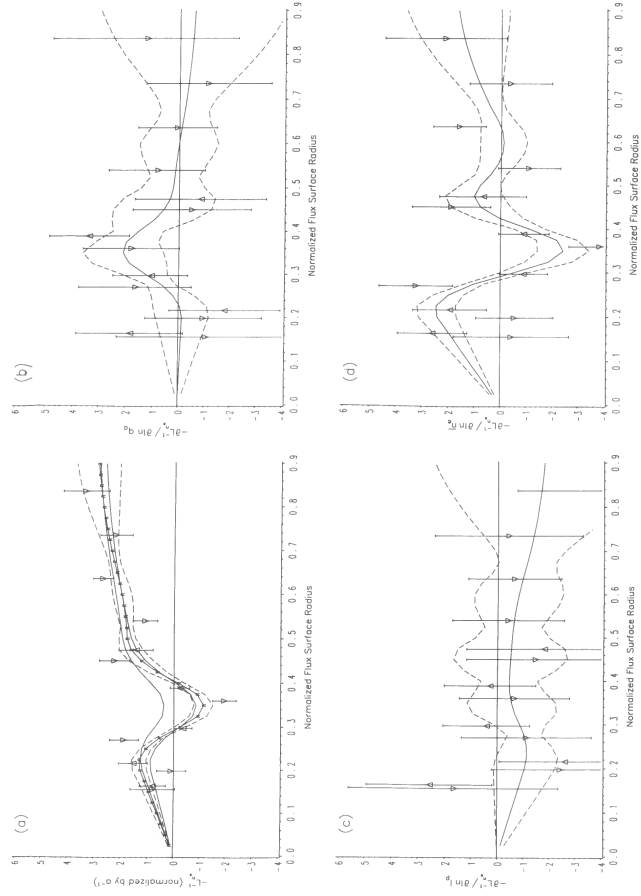

*FIG. 12. Reference negative EFDL profiles for the secondary database $n_e$ and parametric dependence normalized profiles for $q_{sf}$, $I_p$ and $\kappa$.*

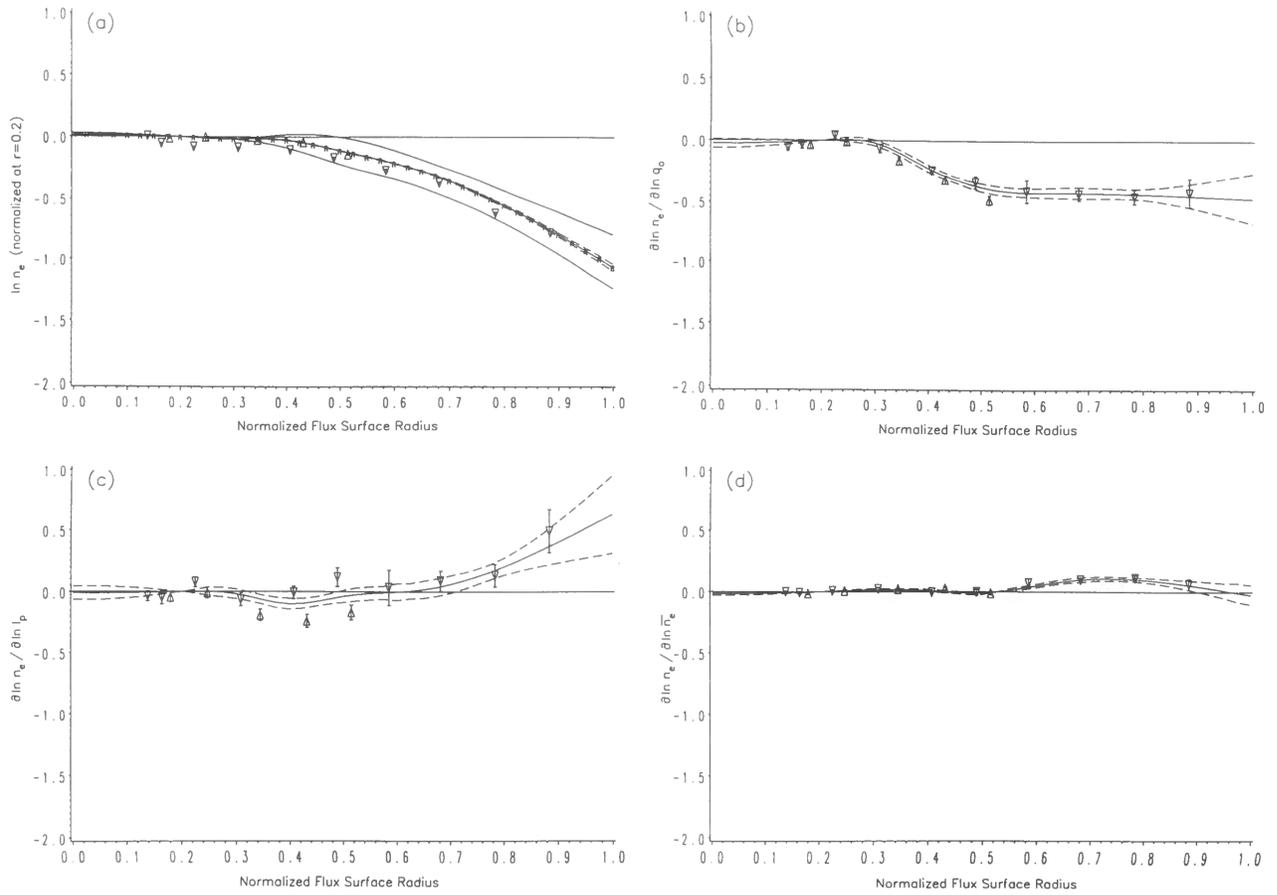

FIG. 13. *Reference* $\ln n_e$ *profiles (normalized at* $r = 0.2$*) for the main database, and parametric dependence profiles for* $q_a$*,* $I_p$ *and* $\bar{n}_r$*.*

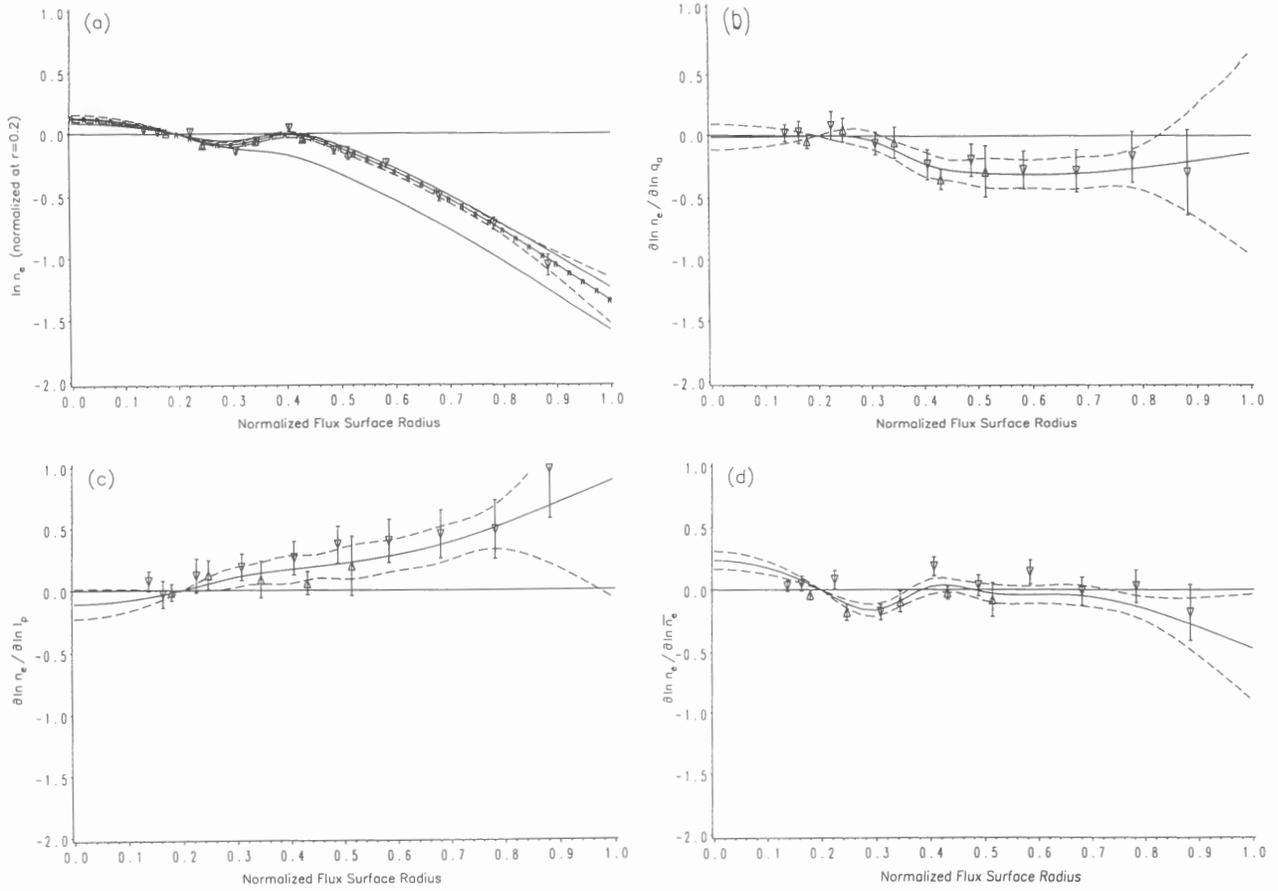

FIG. 14. Reference ln $n_e$ profiles (normalized at r = 0.2) for the secondary database, and parametric dependence profiles for $q_a$, $I_p$ and $\bar{n}_r$.

56

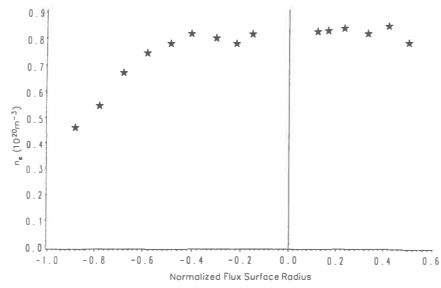

FIG. 15. Sample experimental and predicted density profile with global 95% confidence bands from the main database. Parameters: $q_a = 1.94$, $I_p = 0.432$ MA, $\bar{n}_e = 0.704 \times 10^{20}$ m$^{-3}$.

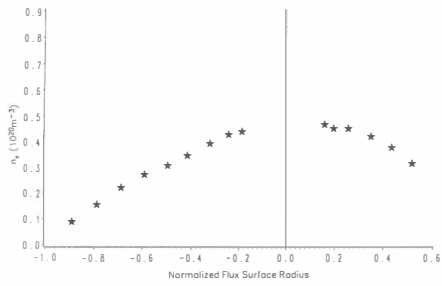

FIG. 16. Sample experimental and predicted density profile with global 95% confidence bands from the secondary database. Parameters: $q_a = 4.01$, $I_p = 0.281$ MA, $\bar{n}_e = 0.299 \times 10^{20}$ m$^{-3}$.

57

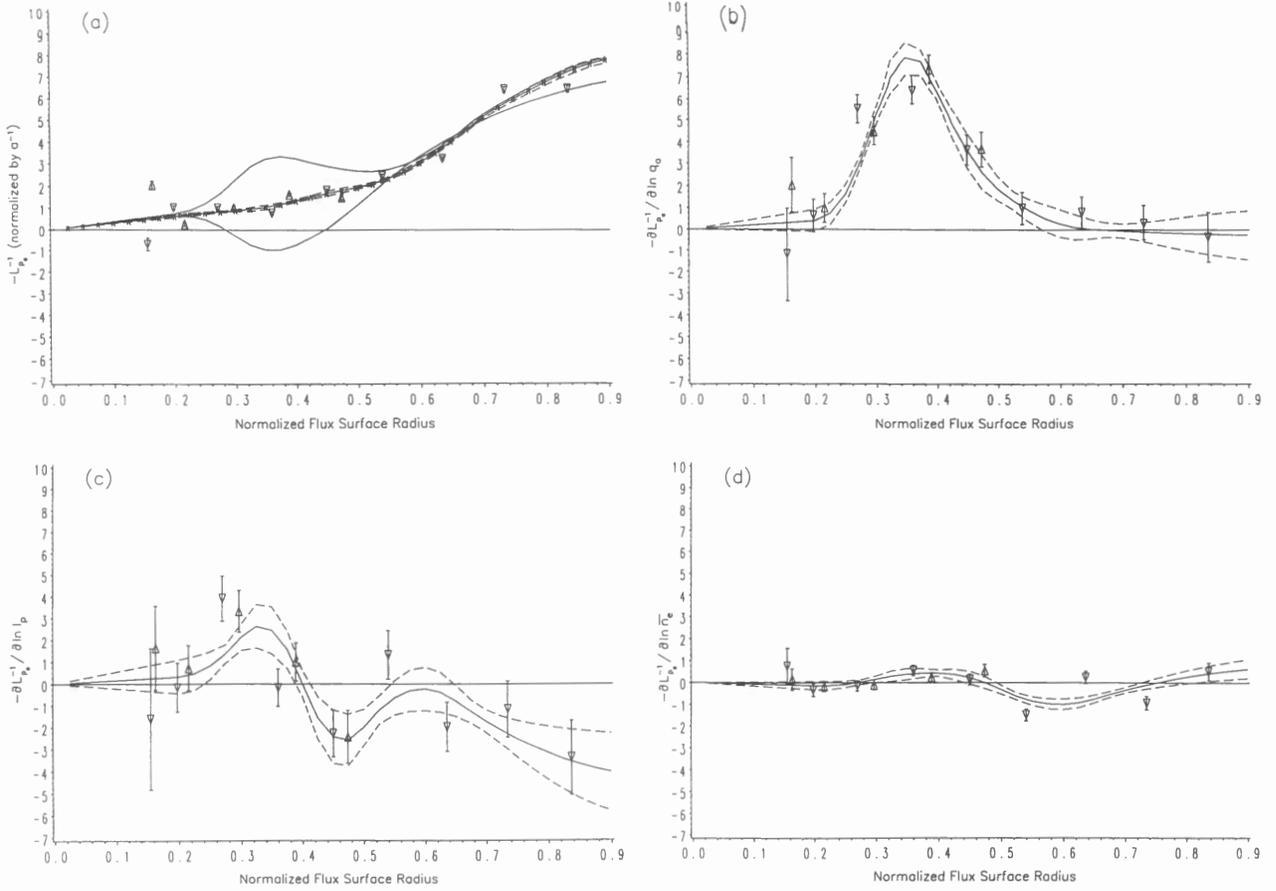

FIG. 17. Reference negative IFOL profiles for the main database $p_e$, and parametric dependence profiles for $q_a$, $I_p$ and $\bar{n}_e$.

58

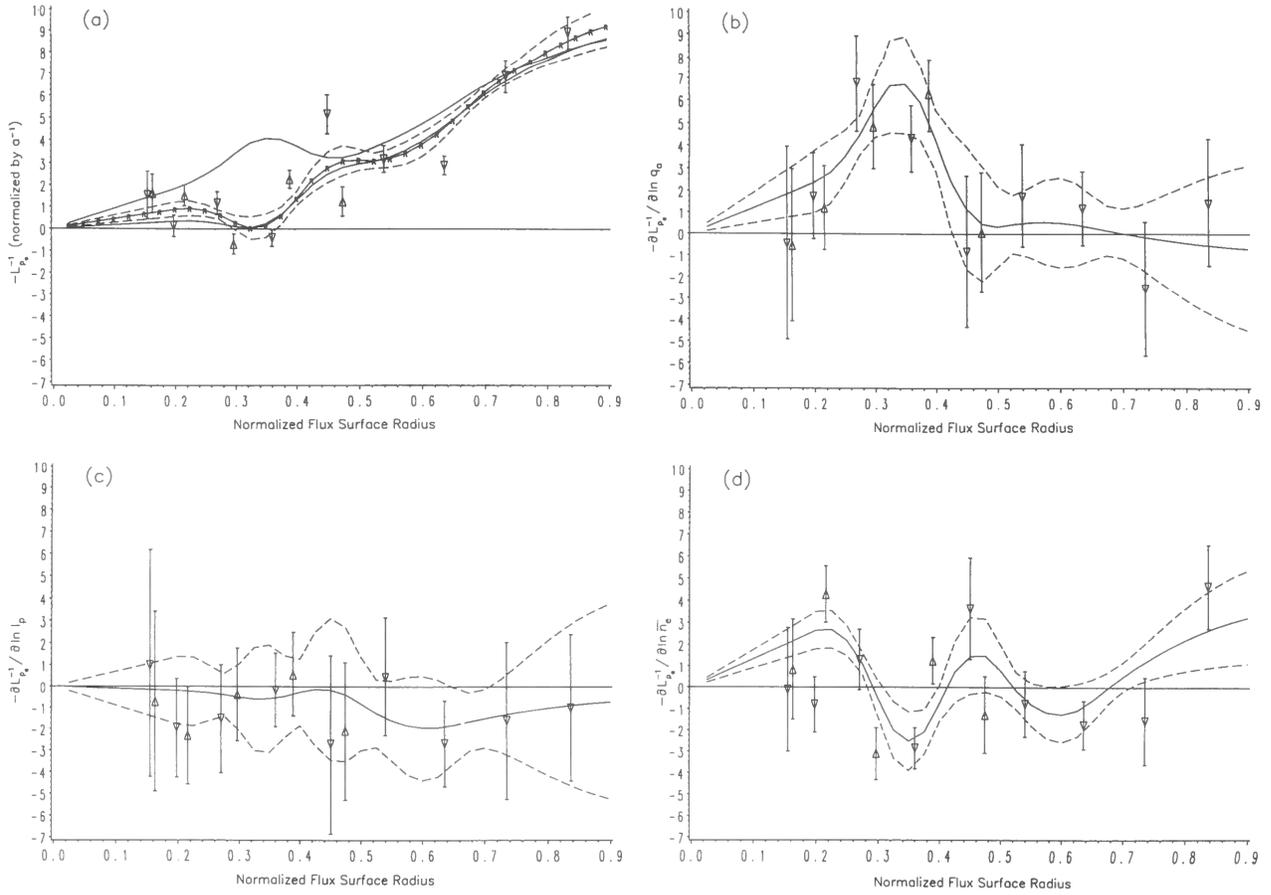

FIG. 18. Reference negative IFOL profiles for the secondry database $p_e$, and parametric dependence profiles for $q_a$, $I_p$ and $\bar{n}_e$.

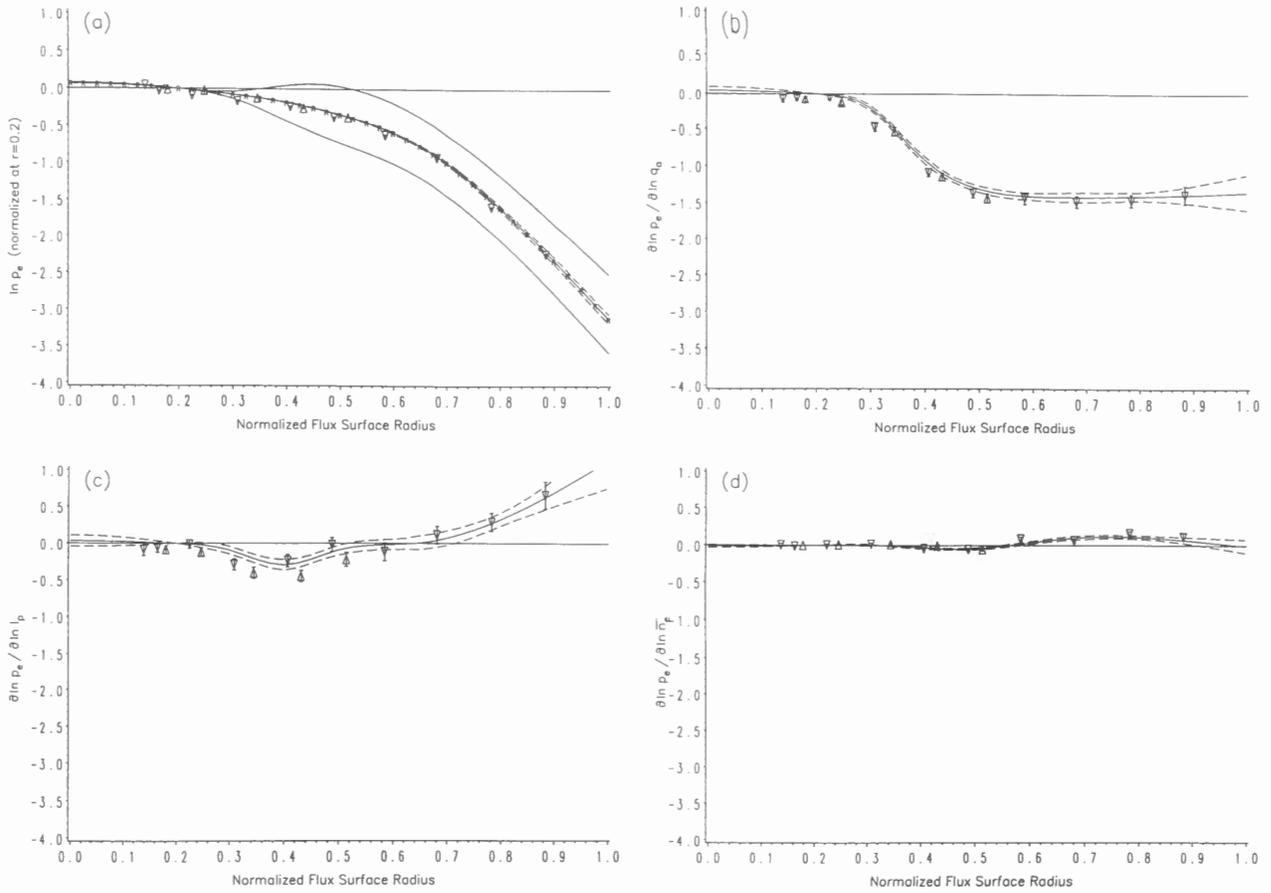

FIG. 19. Reference $\ln p_e$ profiles (normalized at $r = 0.2$) for the main database, and parametric dependence profiles for $q_a$, $I_p$ and $\bar{n}_e$.

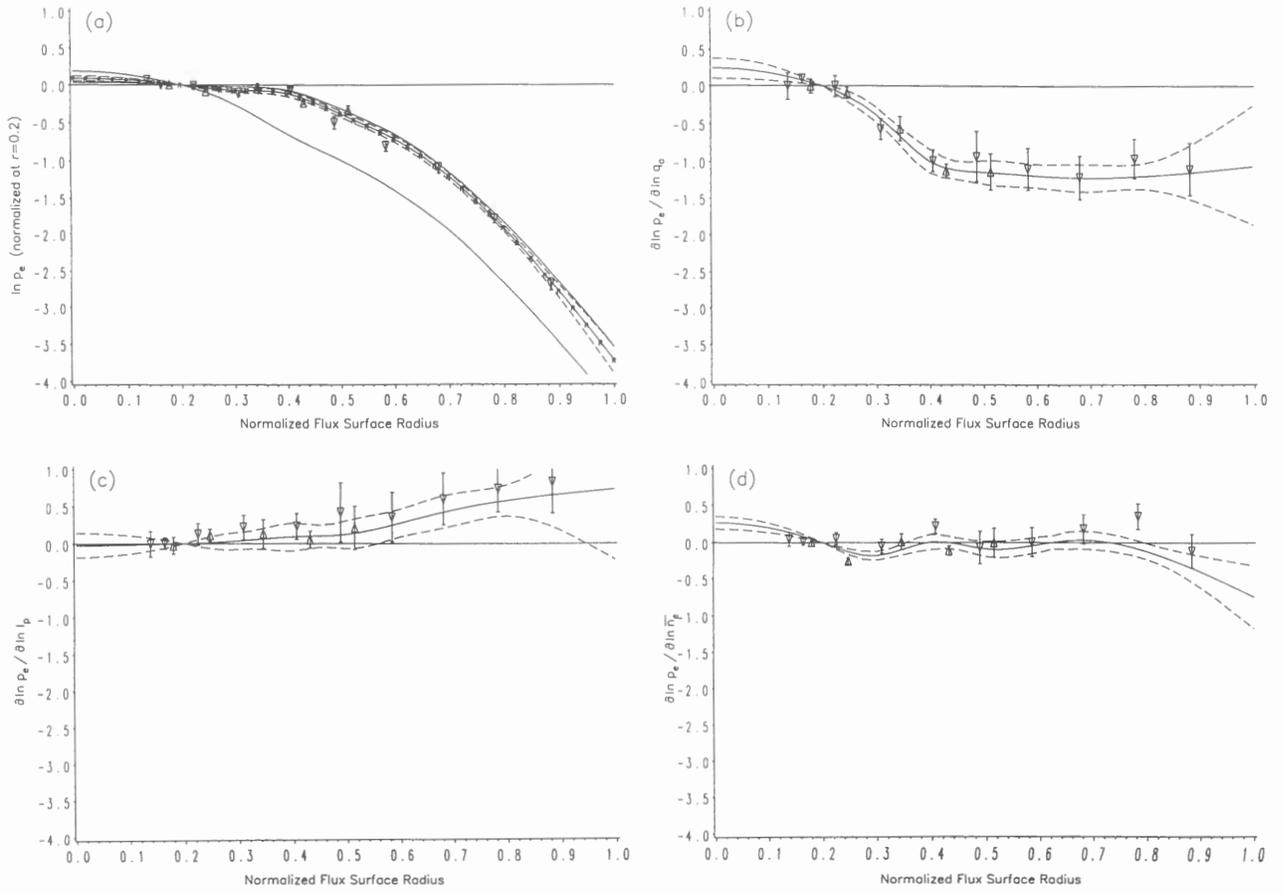

FIG. 20. Reference $\ln p_r$ profiles (normalized at $r = 0.2$) for the secondary database, and parametric dependence profiles for $q_a$, $I_p$ and $\bar{n}_r$.



\end{document}